\documentclass[lettersize,journal]{IEEEtran}

\usepackage{array}
\usepackage[hang]{footmisc}

\usepackage[subrefformat=parens,labelformat=parens,caption=false,font=footnotesize]{subfig}
\usepackage{amsmath, amssymb, amsthm}
\usepackage{cite,cleveref}

\usepackage[nolist,nohyperlinks]{acronym}
\begin{acronym}
    \acro{AP}{access point}
    \acro{ADAS}{advanced driver-assistance systems}
    \acro{CCDF}{complementary cumulative density function}
    \acro{CDF}{cumulative density function}
    \acro{CSMA}{carrier sense multiple access}
    \acro{PDF}{probability density function}
    \acro{PPP}{Poisson point process}
    \acro{BS}{base station}
    \acro{LOS}{line of sight}
    \acro{SINR}{signal-to-interference-plus-noise ratio}
    \acro{SIR}{signal-to-interference ratio}
    \acro{SNR}{signal-to-noise ratio}
    \acro{mm-wave}{millimeter wave}
    \acro{PV}{Poisson-Voronoi}
    \acro{LOS}{line-of-sight}
    \acro{NLOS}{non line-of-sight}
    \acro{PGFL}{probability generating functional}
    \acro{PLCP}{Poisson line Cox process}
    \acro{MBS}{macro base station}
    \acro{MAC}{medium access control}
    \acro{PLT}{Poisson line tessellation}
    \acro{SBS}{small cell base station}
    \acro{RAT}{radio access technique}
    \acro{RCS}{radar cross section}
    \acro{5G}{fifth generation}
    \acro{PLP}{Poisson line process}
    \acro{UE}{user equipment}
    \acro{BPP}{binomial point process}
    \acro{BLP}{binomial line process}
    \acro{BLCP}{binomial line Cox process}
    \acro{RSU}{road-side unit}
    \acro{MD}{meta distribution}
    \acro{SF}{signal fraction}
    \acro{PCP}{Poisson cluster process}
    \acro{UCB}{Upper Confidence Bound}
    \acro{TS}{Thompson Sampling}
    \acro{CSP}{conditional success probability}
    \acro{MAB}{multi-armed bandit}
    \acro{QoS}{quality of service}
    \acro{URLLC}{ultra-reliable low-latency communications}
\end{acronym}

\usepackage{graphicx,wrapfig}

\newtheorem{theorem}{Theorem}[]
\newtheorem{corollary}{Corollary}[]

\newtheorem{lemma}[]{Lemma}
\newtheorem{remark}{Remark}[]

\crefname{figure}{fig.}{Fig.}
\crefname{section}{sec.}{Sec.}

\usepackage{algorithm}
\usepackage{algpseudocode}
\usepackage{scalerel}
\usepackage{multicol}
\usepackage{setspace}
\newtheorem{definition}{Definition}

\usepackage{bbm}
\usepackage[normalem]{ulem}
\usepackage{color}
\usepackage{dsfont}
\usepackage{bm}
\usepackage{setspace}

\newcommand{\cl}[1]{\mathcal{#1}}

\usepackage{caption}
\usepackage{mathtools}
\begin{document}
\title{Statistics of Successive Successful Target Detection in Automotive Radar Networks}

	\author{
 Gourab Ghatak, {\it Member, IEEE}
 \thanks{The author is with the Department of Electrical Engineering at the Indian Institute of Technology (IIT) Delhi, New Delhi, India 110016. Email: gghatak@ee.iitd.ac.in.}
 }

\maketitle

\begin{abstract}
We introduce a novel metric for stochastic geometry based analysis of automotive radar networks called target {\it tracking probability}. Unlike the well-investigated detection probability (often termed as the success or coverage probability in stochastic geometry), the tracking probability characterizes the event of successive successful target detection with a sequence of radar pulses. From a theoretical standpoint, this work adds to the rich repertoire of statistical metrics in stochastic geometry-based wireless network analysis. To optimize the target tracking probability in high interference scenarios, we study a block medium access control (MAC) protocol for the automotive radars to share a common channel and recommend the optimal MAC parameter for a given vehicle and street density. Importantly, we show that the optimal MAC parameter that maximizes the detection probability may not be the one that maximizes the tracking probability. Our research reveals how the tracking event can be naturally mapped to the quality of service (QoS) requirements of latency and reliability for different vehicular technology use-cases. This can enable use-case specific adaptive selection of radar parameters for optimal target tracking.
\end{abstract}


\section{Introduction}
Automotive radars play a critical role in modern \ac{ADAS} and autonomous driving technologies. These systems rely on accurate detection and tracking of moving objects such as vehicles, pedestrians, and other obstacles~\cite{dong2020probabilistic, patole2017automotive}. In particular, radar-based target tracking ensures timely and accurate information about the target’s position, velocity, and trajectory, allowing for safer navigation and collision avoidance~\cite{cao2021automotive}. Given the inherent stochastic nature of the wireless environment, especially in densely populated urban scenarios, developing robust radar tracking algorithms is essential. Radar tracking is a two-step process, comprising both detection and subsequent tracking over multiple time slots with consecutive radar pulses~\cite{howard1990tracking}. The first step involves the successful detection of the target based on the reflected signal \ac{SINR}~\cite{ghatak2022radar}. In a given time slot, the target is successfully detected if the \ac{SINR} exceeds a predefined threshold. Then, a target is considered successfully tracked if it is detected successfully in $\nu$ consecutive slots.

\subsection{Motivation}
The spatial statistics of the detection performance has been thoroughly investigated using tools from stochastic geometry, e.g., see~\cite{al2017stochastic, ghatak2022radar}. In particular, considering the vehicular node locations as points of a point process, researchers have derived the probability of successful detection, the meta-distribution of the reflected signal \ac{SINR}, and consequently, the mean local delay in detection. However, the statistics of consecutive successful detection events is unexplored, which we address in this work. Additionally, our model accounts for street intersections and the corresponding incoming traffic in analyzing the system performance. This is critical for developing automotive radar aided safety applications but has limited literature.

\subsection{Related Work}
In recent years, studies based on stochastic geometry have gained traction to model and analyze the statistics of automotive radar interference while addressing the variability in radar deployments, e.g., see~\cite{al2017stochastic, munari2018stochastic, fang2020stochastic, ghatak2022radar, ghatak2021fine}. These models are crucial for accurately evaluating radar system performance and determining the impact of interference mitigation strategies. For example, the study in~\cite{al2017stochastic} modeled the distribution of automotive radars using a \ac{PPP} and estimated the average \ac{SIR} for radar detection. Munari \textit{et al.}~\cite{munari2018stochastic} employed the strongest interferer approximation to assess radar detection range and false alarm rates. Additionally, the work in~\cite{fang2020stochastic} examined radar detection probabilities by incorporating fluctuating radar cross-sections (RCS) based on Swerling-I and Chi-square models, offering more precise and intuitive outcomes compared to constant RCS models. A fine grained analysis of the impact of interference is studied in \cite{ghatak2022radar} where the authors derived not only the \ac{MD} of the \ac{SINR} but also employed it to study the mean local delay in radar detection. Furthermore, they proposed an online algorithm to control the channel access by individual radars. Radar \ac{MAC} protocols such as ALOHA and \ac{CSMA} have also been investigated in \cite{haritha2021slotted, sukumaran2022slotted}. Furthermore, asynchronous non-cooperative protocols for MAC was studied in~\cite{9399786}. By enhancing these models, we can improve performance evaluations, leading to the development of more optimized solutions. One of the primary advantages of using stochastic geometry frameworks is the ability to explore various spatial scenarios with significantly reduced computational resources and time compared to traditional system-level Monte Carlo simulations or real-world measurement data collection. \textcolor{blue}{Additionally, it must be noted that while real-world radar datasets and high fidelity system-level simulations are valuable for empirical benchmarking, they remain difficult to scale or generalize due to their inherent limitations in capturing large-scale interference and deployment randomness. In contrast, our stochastic geometry framework balances realism and analytical tractability, enabling scalable and generalizable performance analysis.}

While these studies predominantly focus on two-lane scenarios, the authors in~\cite{chu2020interference} extended the analysis to multi-lane scenarios by utilizing a marked point process model to characterize radar interference. Furthermore, more recent work has incorporated Matern Hard-Core Processes (MHCP)~\cite{mishra2020stochastic,kui2021interference}, which account for vehicle dimensions and the required road space, providing a more realistic interference model for automotive radar systems. However, in order to gain a more accurate view of the network, a complete two dimensional spatial model is paramount, which none of the above works address. In this regard, novel line processes have been proposed to model and study structure of streets and vehicular networks~\cite{shah2024binomial, shah2024modeling}. In this work, we consider the \ac{PLCP} model for emulating the streets in order to accurately incorporate the impact of radar beamwidth (e.g., see~\cite{nabil2024beamwidth} for factors impacting beamwidth selection in such networks) and the incoming traffic at intersections.

More importantly, all the above studies either focus on one shot detection, i.e., the event of successful target detection in one radar pulse transmission attempt, or characterize the mean local delay, i.e., the number of transmission attempts for first successful detection. Albeit useful, these metrics cannot estimate the performance of target tracking. To address this, we derive the distribution of run-length of successes and accordingly, characterize the statistics of successive successful detection, or tracking\footnote{From a control systems perspective, the run-length distribution was also characterized in \cite{ghatak2024RAWNET}, however, there the authors did not study the same in a \ac{PLCP} model and did not employ the same to study an automotive radar network, which presents its own challenges.}.

\subsection{Contributions}
The main contributions of this work are as follows.
\begin{itemize}
    \item We define and derive a new metric called the tracking probability to study automotive radar networks. It characterizes the event that the ego radar is able to detect the target vehicle successfully in successive time slots. Although relevant for reliability constrained applications, this metric has not been studied yet in the context of stochastic geometry. The key technical challenge to derive this is to handle the temporal correlation in run-length of successes in a fixed time window, which is overcome using de Moivre's theorem.
    \item Then, we introduce a novel block ALOHA-based radar \ac{MAC} protocol where active radars access the channel in a block of $T$ consecutive slots with block probability $\delta$. Unlike the slot-based ALOHA protocol typically assumed in automotive radar networks (e.g., see \cite{ghatak2022radar, munari2018stochastic}), the block ALOHA enables an enhanced radar tracking probability. Then we derive the detection probability of the ego radar, followed by the conditional distribution of the number of successive slots in which the reflected signal \ac{SINR} exceeds a detection threshold. Furthermore, we derive the expected number of such successful tracking events in the block. As deriving the distribution of the conditional tracking probability is infeasible, we analyze its first moment, which denotes the probability of successful target tracking across all network realizations.
    \item Finally, we study the optimization of the tracking probability of the ego radar with respect to the channel access probability and the radar beamwidth. This enables us to derive insights into the design of adaptive cognitive radars, specifically, how the radar parameters may be adapted in order to enable better tracking based on the street geometry, vehicle density, and use-case \ac{QoS} requirements of latency and reliability. We show that the latency deadline of a service can be partitioned into access blocks consisting of multiple detection slots, each of which corresponds to a radar pulse width. We study the fundamental trade-off in this frame design between the block length and the number of blocks and highlight its role in maximizing successful radar tracking.
\end{itemize}

 \subsection{Key Technical Challenges:} 
    {Most of the stochastic-geometry based analysis of automotive radar networks were focused on a single or double lane highway scenario with a simple one dimensional PPP to emulate the location of the interferers (e.g., see \cite{al2017stochastic}). Although the PLCP is a well-studied component in stochastic-geometry, most of its characterization has focused on the perspective of cellular networks. On the contrary, due to the consideration of automotive radar networks in this paper, the interfering set of the vehicles is formed by the intersection of the PLCP with a double conic set. This makes the interfering set a non-stationary point process, which requires the derivation of the conditional interference lengths for a given intersection distance. The challenge is compounded by the fact that in contrast to the typical one-shot detection probability, due to the correlation of run-lengths, the tracking characterization is a challenging task.
    }

The rest of the paper is organized as follows. In Section~\ref{sec:SM} we describe the system model, characterize the interference set, and outline the key definitions of radar detection and radar tracking. Our main results are presented in Section~\ref{sec:SRT} where we derive the statistics of radar tracking. Then, in Section~\ref{sec:OFD}, based on the derived framework, we explore \ac{QoS}-Aware optimal frame design. Numerical results to highlight the salient features and insights in our framework are presented in Section~\ref{sec:NRD}. Finally, the paper concludes in Section~\ref{sec:Con}.

\section{System Model and Key Definitions}
\begin{table*}[t]
\small{
\centering
\centering
\begin{tabular}{|l|l|l|l|}
\hline
Parameter name & Symbol & Units \\ \hline
Transmit power & P & dBW\\ \hline
Target distance & R & m\\ \hline
Radar cross section & $\sigma_{\rm c}$ & m$^2$\\ \hline
Path loss exponent & $\alpha$ & -\\ \hline
Noise power & $N_0$ & dBW\\ \hline
SINR Threshold & $\gamma$ & dB\\ \hline
SINR & $\xi$ & dB\\ \hline
Tracking length & $L$ & slots \\ \hline
Reliability (minimum tracking run length) & $\nu$ & slots \\ \hline
Block length & T & slots \\ \hline
Minimum and maximum interference region in an intersecting street & $a$ and $b$ & m \\ \hline
\end{tabular}
\caption{System parameters and definitions.}
\label{table:parameters}}
\end{table*}
\label{sec:SM}
\subsection{Street Geometry and Vehicle Locations}
We consider a network of streets modeled as a homogeneous \ac{PLP}, $\mathcal{P} = \{L_1, L_2, \dots\}$~\cite{ghatak2019small}. A line $L_i \in \mathcal{P}$ corresponds to a tuple $(\theta_i, r_i)$, where $r_i$ is the length of the normal to $L_i$ from the origin and $\theta_i$ is the angle between the normal and the $x$-axis. The parameters $\mathbf{q}_i := (\theta_i,r_i)$ reside in the representation space $\mathcal{D} \equiv [0,2\pi) \times (0,\infty)$. The number of parameter points, $I$, in any $S \subset \mathcal{D}$ follows the Poisson distribution with parameter $\lambda_{\rm L}|S|$, where $|S|$ represents the Lebesgue measure of $S$. The locations of the vehicles with mounted radars on the street $L_{i}$ follow a one-dimensional \ac{PPP}, $\Phi_{{\rm L}_i}$, with intensity $\lambda$, that represents the vehicular density. The \ac{PPP} of the vehicle locations on any street is independent of the corresponding distributions on the other streets. Hence, the complete distribution of the vehicles is a homogeneous \ac{PLCP} $\Phi$, on the domain $\mathcal{P}$ where $\Phi = \bigcup_{{\rm L}_i \in \mathcal{P}} \Phi_{{\rm L}_i}$. Our analysis is carried out from the perspective of an ego radar located at the origin of the two-dimensional plane. We model the ego radar as the typical point of $\Phi$.
{\color{blue} It is important of highlight that while this assumption facilitates tractable analysis, it does not capture potential correlations in the spatial deployment of radars along structured line processes (e.g., roadways or corridors). An alternative approach is to employ the \ac{BLCP} or inhomogeneous \ac{PLCP}, which allows for more realistic modeling of radar interferers distributed along random lines. Incorporating such inhomogeneity is left outside the present scope and is discussed in detail in the numerical results section.} In particular, let us define
\begin{align}
    \Phi^o\triangleq(\Phi\mid o\in\Phi), \nonumber
\end{align}
for any point $o$, and note that under expectation over $\Phi^o$, $o$ becomes the typical point of the process $\Phi$~\cite{chiu2013stochastic}. Since by construction, $\Phi$ is a stationary process, let us consider $o$ to be the origin of $\mathbb{R}^2$. Under Palm conditioning, a line $L_0$ passes through the origin almost surely, which without loss of generality, we consider to be the $y-$axis. The ego radar experiences interference if and only if both the ego radar and interfering radar fall into each other's radar sectors simultaneously. To model the interference from the vehicles traveling to the opposite side as the ego radar, we consider on line $L_0$ the location of vehicles follows a one-dimensional \ac{PPP} having the same intensity as the remaining lines. Therefore, the overall spatial stochastic process of potential interfering radars, $\Phi_{{\rm P}_0}$, as per the Palm conditioning, is the superposition of $\Phi$ and an independent 1D \ac{PPP} on line $L_0$, i.e., $\Phi_{{\rm P}_0} = \Phi \cup \Phi_{L_0}$.

\subsection{Signal and Channel Model}
The radars have a half-power beamwidth of $\Omega$, and the target is assumed to be located at a distance $R$ from the radar on the same street within the main beam of the ego radar. Let the transmit power be $P$ and the path-loss exponent be $\alpha$. {The target is assumed to have a Swerling-I fluctuating radar cross-section, $\sigma_{\rm c}$, with a mean of $\bar{\sigma}$~\cite{shnidman2003expanded}. The fluctuating radar cross section is due to multiple factors such as variation in the target aspect angle, multipath effects, and different target material composition. The Swerling-I is a popular model for emulating such variations. Although our work can be easily extended to incorporate other models by appropriately modifying the distribution, we assume the simple Swerling-I model in order to not divert focus away from the main aspect of our work - statistics  of successive successful detection.} Let $G_t$ be the gain of the transmitting antenna and $A_{\rm e}$ the effective area of the receiving antenna aperture. {Then, the reflected power from the target at the ego radar is $S = \chi\sigma_{\rm c} P R^{-2\alpha}$, where $\chi = \frac{G_{\rm t}}{(4\pi)^2}A_{\rm e}$ is the product of the antenna gain and the fractional aperture of the receiver}. Let the coordinates of any interfering radar be denoted by $\mathbf{w} = (x,y)$, such that $x\cos\theta_i + y\sin\theta_i = r_i$ for $(\theta_i,r_i) \in \mathcal{D}_{\rm P}$. The interfering signals undergo independent multi-path Rayleigh fading with parameter 1~\cite{ghatak2022radar}. Thus, the interference at the ego radar due to an interfering radar located at $w$ is $\mathbf{I} = P \gamma h_{\mathbf{w}} ||\mathbf{w}||^{-\alpha}$, where $h_{\mathbf{w}}$ is the fading power. Now, assuming that all the automotive radars share the same power and gain characteristics, the \ac{SINR} at the ego radar is
\begin{align}
    \xi(R) = \frac{\chi \sigma_{\rm c} P R^{-2\alpha}}{N_0 + \sum_{\mathbf{w} \in \Phi_{\rm I}}4\pi \chi P h_{\mathbf{w}} ||\mathbf{w}||^{-\alpha}},
\end{align}
where where {$N_0$ is the noise power, i.e., the noise density multiplied by the transmission bandwidth} and $\Phi_{\rm I} \subset \Phi_{{\rm P}_0}$ is the interfering set, defined and derived in the next subsection.
\subsection{Interfering Set and Distance}
The orientation of the vehicles on the street depends on the generating angle of the street and the direction of their movement. Thus, the boresight direction of any radar on the line $L_i$ is given by the two unit vectors: $\mathbf{z}_i$ and $-\mathbf{z}_i$, where $\mathbf{z}_i = (-\sin\theta_i, \cos\theta_i)^T$. The parameters $(\theta_i,r_i)$, determines the direction of the radar sector on the street $L_i$. Any point $(p,q) \in \mathbb{R}^2$ lies in the sector if the angle made by the displacement vector between $(p,q)$ and $(x, y)$ and the boresight direction is larger than $\cos \Omega$. Thus, the sector of a radar located at $(x,y)$ is either $\mathcal{R}^{+}_{(x,y)}$ or $\mathcal{R}^{-}_{(x,y), k}$ based on its direction, mathematically defined below.
\begin{definition}{\cite{shah2024modeling}}
\label{def:def_1}
The radar sector of a vehicle located at $(x,y) \in \Phi_{{\rm P}_0}$ for boresight direction $\mathbf{z}_i$ is
\begin{align*}
    \mathcal{R}^{+}_{(x,y)} &= \Bigg\{(p,q) \in \mathbb{R}^2 \colon \frac{\big((p,q) - (x, y)\big) \cdot \mathbf{z}_i}{||(p,q) - (x, y)||} > \cos \Omega, \\
    &\hspace*{3.5cm} ||(p,q) - (x, y)|| \leq R \Bigg\}\\
    &\hspace*{-1.25cm}= \Bigg\{(p,q) \in \mathbb{R}^2 \colon \frac{\cos\theta_i\left(q - y\right) - \sin\theta_i\left(p -  x\right)}{\sqrt{(p -  x)^2 + (q - y)^2}} > \cos \Omega,\\
    &\hspace*{2.8cm} \sqrt{(p -  x)^2 + (q - y)^2} \leq R \Bigg\}.
\end{align*}
And for boresight direction $-\mathbf{z}_i$,
\begin{align*}
    \mathcal{R}^{-}_{(x,y)} &= \\
    &\hspace*{-1cm}\Bigg\{(p,q) \in \mathbb{R}^2 \colon \frac{\sin\theta_i\left(p -  x\right) - \cos\theta_i\left(q - y\right)}{\sqrt{(p -x)^2 + (q - y)^2}} > \cos \Omega,\\
    &\hspace*{2.3cm}; \sqrt{(p -  x)^2 + (q - y)^2} \leq R \Bigg\}.
\end{align*}
\end{definition}
\begin{figure}
    \centering
    \includegraphics[width=0.5\linewidth]{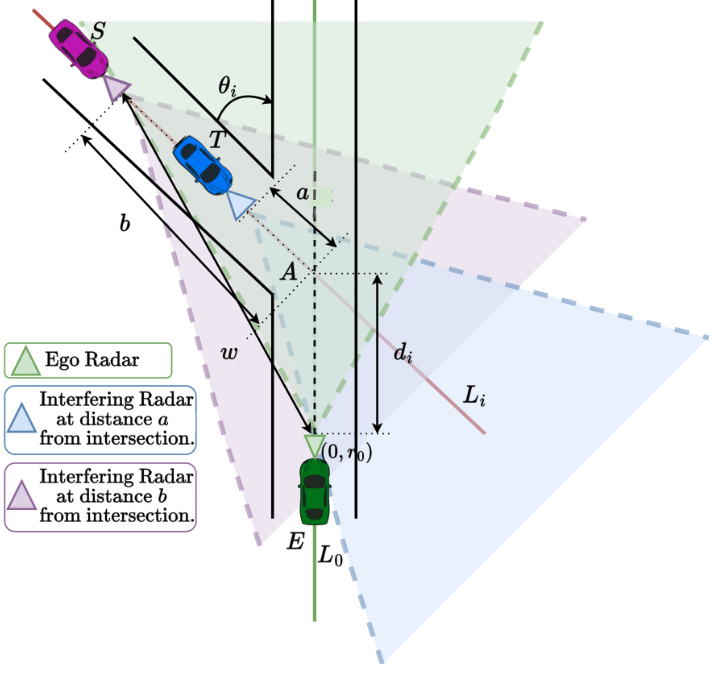}
    \caption{Illustration of the interfering ranges in an intersecting street in $L_0$.}
    \label{fig:illustration}
\end{figure}
The ego radar sector corresponds to the line $(\theta_i,r_i) = (0,0)$ and the location $(x,y) = (0,0)$. If ego radar is located within the radar sector of the radar present at $(x,y)$ i.e., $\mathbf{1} \left((0,r) \in \mathcal{R}^{+}_{(x,y)}\right) + \mathbf{1} \left((0,r) \in \mathcal{R}^{-}_{(x,y)}\right)$ evaluates to one, and simultaneously the automotive radar located at $(x,y)$ is located in the interior region of ego radar sector i.e., $\mathbf{1} \left((x,y) \in \mathcal{R}^{+}_{(0,r)}\right) + \mathbf{1} \left((x,y) \in \mathcal{R}^{-}_{(0,r)}\right)$ evaluates to one, then radar at $(x,y)$ will be an element of $\Phi_{\rm I}$. {Mathematically, this thinning procedure is represented as
\begin{align}
    \Phi_{\rm I} = \Phi_{{\rm P}_0} \cap \mathcal{DC}(r),
\end{align}
where $\mathcal{DC}(r)$ is the double-conic intersection region represented as:
\begin{align}
\mathcal{DC}(r) = &\{(x,y): \left[\mathbf{1} \left((0,r) \in \mathcal{R}^{+}_{(x,y)}\right) + \mathbf{1} \left((0,r) \in \mathcal{R}^{-}_{(x,y)}\right)\right] \nonumber \\
&\cdot \left[\mathbf{1} \left((x,y) \in \mathcal{R}^{+}_{(0,r)}\right) + \mathbf{1} \left((x,y) \in \mathcal{R}^{-}_{(0,r)}\right)\right] = 1\}.
\label{eq:mutual_interference}
\end{align}
For the ego-radar, $\Phi_{\rm I}$ represents the set of vehicles which lie in the ego-radar's beam while simultaneously the ego-radar lies in their beam, thereby causing mutual interference.} {This is further illustrated in Fig.~\ref{fig:illustration}, where the ego radar is shown in green. An intersecting street $L_i$ crosses $L_0$ at a distance $d_i$ from the ego radar. As a result, the variables $a$ and $b$ are defined from the point of intersection of $L_i$ and $L_0$. In order to calculate $w$, we further need the angle of intersection, which is denoted by $\theta_i$ in the figure. Two extremes of interfering radar locations are shown with blue (at location $T$) and magenta (in location $S$), respectively. Now, on this street $L_i$, any vehicle if present between $S$ and $T$ will create an interference to the ego-radar. Any vehicle which is present at a distance greater than $b$ (i.e., behind $S$) will not create interference since it will not be present in the ego-radar beam. On the contrary, any radar present at a distance lower than $a$ (i.e., in front of $T$) also will not cause interference to the ego radar since the ego-radar will not be present in the beam of such a vehicle. For all such streets intersecting $L_0$, the interfering region needs to be characterized for such $a$ and $b$ limits. This is formally presented in the following result.}
\begin{lemma}
\label{le:lemma1}
For line $L$ characterized by the PLP point $(r, \theta)$ where $0 < \theta \leq \frac{\pi}{2}$, that intersects $L_0$ at a distance $d$ from the ego radar, the minimum ($a$) and the maximum ($b$) distances of the interfering radars from the point of intersection are
\begin{align}
  a &= \frac{d\sin \left(\theta - \Omega/2\right)}{\sin \Omega/2} \nonumber \\
  b &= \frac{d\sin \Omega/2}{\sin \left(\theta - \Omega/2\right)}. \nonumber
\end{align}
\end{lemma}
\begin{IEEEproof}
Please see Appendix~\ref{app:lemma1}.
\end{IEEEproof}
With this characterization we note that $\Phi_{\rm I} = \cup_{i \in \mathcal{P}} \left( \Phi_{{\rm P}_0} \cap (a_i, b_i) \mathbf{z}_i \right)$. It must be noted that here we have made an abuse of notation by referring to the \ac{PLCP} as a subset of the \ac{PLP} thus allowing for the intersection of the elements of the two sets. This is well-defined iff the PLP itself is defined as a subset of $\mathbb{R}^2$ and not merely a set of lines. However, this notation allows us to simplify the expressions and does not cause any deviation from the correctness of the results. Based on this interfering set, we study the tracking performance of a radar access protocol defined next.

\subsection{Radar \ac{MAC} Protocol}
\begin{figure}
    \centering
    \includegraphics[width=0.75\linewidth]{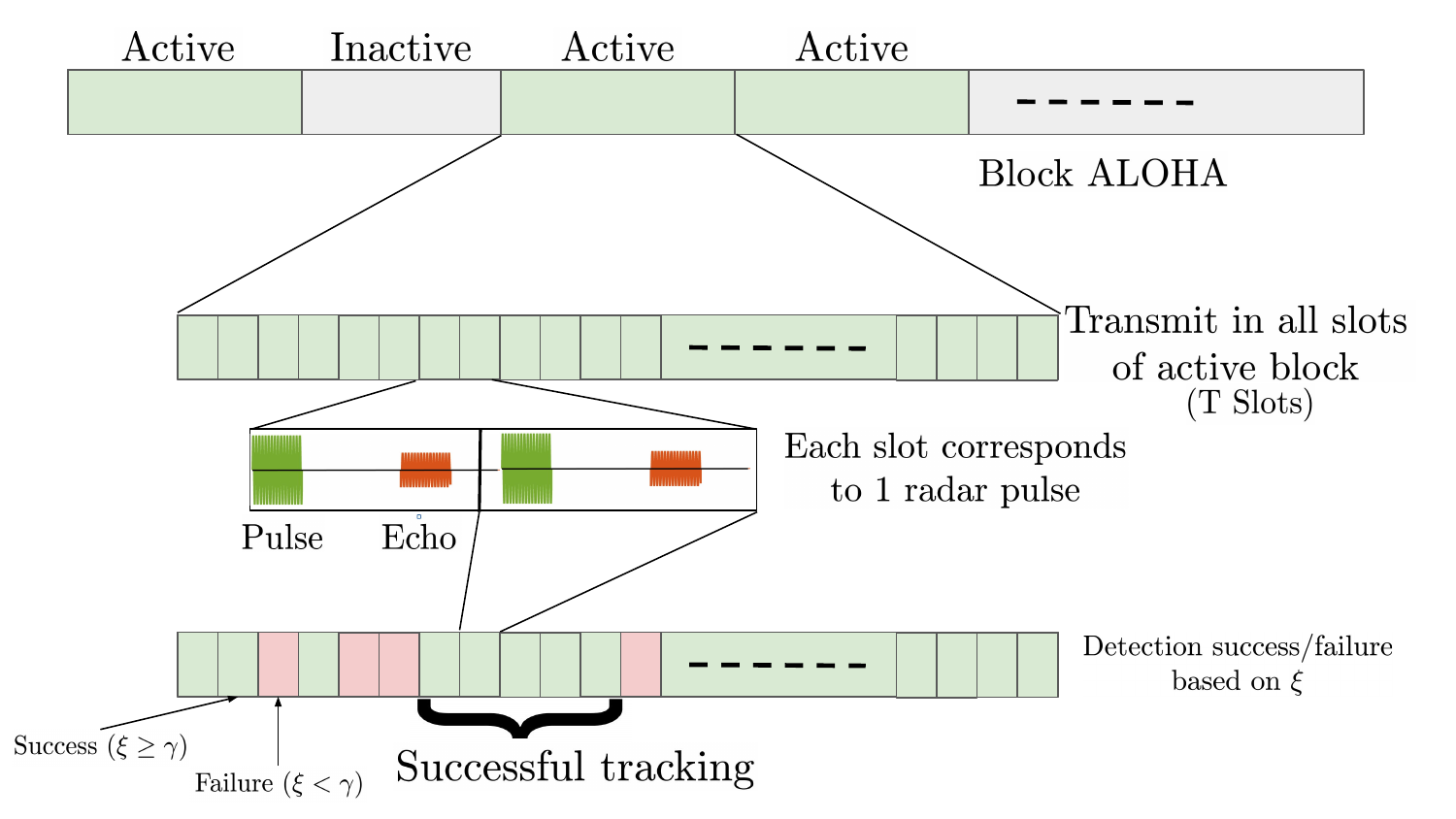}
    \caption{Illustration of the frame design and the MAC protocol from the perspective of the ego-radar.}
    \label{fig:system}
\end{figure}

In dense vehicular networks, in particular where the radar and communication bands are shared, frequent radar signals may degrade the overall detection as well as the communication performance. For this, we propose a tunable block ALOHA protocol for shared radar channel access. {\color{blue} We adopt a centralized assumption for determining optimal ALOHA parameters, wherein the network computes and disseminates the settings rather than vehicles making decentralized decisions. This modeling choice is justified by the stringent reliability and latency requirements of automotive radar systems, where distributed contention-based optimization would introduce instability and risk violating the end-to-end latency budgets mandated by V2X safety applications (e.g., $<5$ ms in 3GPP Rel-18~\cite{3gpp_rel18_v2x}). As vehicles increasingly operate as 5G/6G UEs, network entities such as gNBs or RSUs can act as controllers that compute globally optimal radar access parameters and broadcast them via system information blocks with negligible overhead, thereby supporting \ac{URLLC}. This network-assisted dissemination is consistent with ongoing NR-V2X standardization efforts~\cite{3gpp_rel18_v2x}, whereas no standardized MAC protocol currently exists for automotive radars, with existing solutions remaining vendor-specific and at the pre-standardization research stage.} {Since our focus is on the characterization of the tracking probability, we assume a simplistic block ALOHA protocol. Other MAC schemes such as CSMA would require the development of hard-core point processes, e.g., see~\cite{alfano2012new}, which we will treat in a future work.}

Let the time frames be divided into blocks of $T$ slots each. In each block, only a subset of the vehicles are allowed to activate their radars. The activation set follows the classical ALOHA protocol with parameter $\delta \in [0,1]$. Thus, each vehicle in $\Phi_{\rm P}$ operates in a block with probability $\delta$. Within the block {in which it is active}, it attempts to detect the target in each of the $T$ slots. We present our analysis from the perspective of a single such block in which the ego radar is active. Later, we discuss the intricacies of designing the block length in Section IV. Here we define the key metrics that we statistically characterize in the next section.
\begin{definition}[Successful target detection]
\label{def:det}
We define the target to be successfully detected in a slot if $\xi \geq \gamma$ for some threshold $\gamma$.    
\end{definition}
Let $L$ denote the maximum run length of detection successes in a block for the ego radar, i.e., it is the largest number $L \in [0, T]$, such that the ego radar experiences $L$ consecutive successful target detection in consecutive. We call $L$ as the tracking length, which is a random variable.

\begin{definition}[Successful target tracking]
\label{def:track}
We define the target to be successfully tracked in a block of length $T$ slots if the ego-radar successfully detects the target in $\nu$ consecutive slots, i.e., the event $L \geq \nu$, for some $\nu \leq T$.
\end{definition}
{We refer to $\nu$ as the reliability parameter of the radar application. The higher the value of $\nu$, more stringent is the requirement from the radar system. The overall protocol is illustrated in Fig.~\ref{fig:system}.  Due to the block ALOHA access policy, the radar gets access in only a fraction of the blocks. In the active blocks, the radar attempts to detect the target in each of the $T$ slots, i.e., it transmits on all the slots of the active block. Each slot corresponds to one radar pulse as shown. The success of a slot is defined in Def.~\ref{def:det}, while the target is termed to be successfully tracked if at least $\nu$ consecutive slots are successful. An illustration with $\nu = 5$ is shown.}

{The key challenge in characterizing the above is the fact that the bursts are not independent of each other. To illustrate, consider the event $E_{[t_1,t_2]}$ that all the slots from $t_1$ to $t_2$ of the block were successful (i.e., the target was successfully detected). Let $\bar{E}_{[t_1,t_2]}$ represent the complement of the event. Naturally, assuming $\nu = 4$, $\mathbb{P}\left(E_{[2,5]} | \bar{E}_{[1,4]}\right) = 0 \neq \mathbb{P}\left(E_{[2,5]}\right)$. Indeed given that the first four slots have failed, the probability that the slots 2 to 5 has a successful run length is 0 due to the slots $2, 3,$ and $4$ being common to both the events.}
\section{Statistics of Radar Tracking}
\label{sec:SRT}
In order to characterize the radar tracking, first we derive the probability of a successful detection event in one slot given that the ego radar is active in the block.
\begin{lemma}
The conditional probability of successful target detection in a slot by the ego radar given that it is active in the block is given by
    \begin{align}
        P_{\rm d} =  \exp{\left(\frac{-\gamma}{\xi_0 R^{-2\alpha}}\right)} \left( \prod_{{\bf w} \in \phi^{\prime}} \frac{1}{1+\frac{\gamma d_{\bf w}^{-\alpha}}{R^{-2\alpha}}} \right), \label{eq:cond_succ}
    \end{align}
    where $\xi_0 = \frac{P}{N_0}$ is the transmit \ac{SNR} and $\phi'$ is an instance of $\Phi'$ which is a thinned version of $\Phi_{\rm I}$ where the \ac{PPP} on each line has an intensity $\delta \lambda$.
    \label{lemma_csp}
\end{lemma}
\begin{IEEEproof}
Please see Appendix~\ref{app:csp}.    
\end{IEEEproof}
It is worth to note that due to the conditioning on $\Phi$, $P_{\rm d}$ is a random variable defined on a suitable sigma field where the counting measure $\Phi$ is defined. Indeed, conditioned on $\Phi$, the above success event is an independent random variable across the time slots of a block when the ego radar is active. It is precisely this random variable that we characterize in this work. The probability $P_{\rm d}$ changes across blocks as the set of active radars changes.
{\color{blue}
\begin{remark}
The two constituent terms of the conditional detection probability correspond to the impact of the noise and the impact of the interference on the detection performance of the ego radar. Naturally, for a noise-limited regime, the second term of the above should be considered 1, while for an interference-dominated regime, the first term should be considered 1. This is mathematically stated as
\begin{align}
    P^{\text{Noise limited}}_{\rm d} &= {\exp{\left(\frac{-\gamma}{\xi_0 R^{-2\alpha}}\right)}} \nonumber \\
    P^{\text{Interference dominated}}_{\rm d} &=  \prod_{{\bf w} \in \phi^{\prime}} \frac{1}{1+\frac{\gamma d_{\bf w}^{-\alpha}}{R^{-2\alpha}}}
\end{align}
\end{remark}
}
\begin{remark}
The conditional success probability in \eqref{eq:cond_succ} is different from that derived in~\cite{ghatak2022radar, haritha2021slotted, sukumaran2022slotted}. In particular, the above works consider random access in each slot and hence, there the transmission state of each radar needs to be averaged out with respect to the ALOHA parameter. However, in our case, once the transmission states of the radars are determined at the beginning of the block, the same remains fixed throughout all the slots of the block. Hence, the impact of the ALOHA only appears in the set over which the product of the point functions is evaluated in \eqref{eq:cond_succ}.
\end{remark}
{\color{blue}
In the above, we assumed Rayleigh fading for the interfering signals. The framework can be extended to account for log-normal shadowing by introducing a multiplicative shadowing factor $X_w \sim \text{Lognormal}(0,\sigma^2)$ on each interfering link. In this case, the interference from radar $w$ becomes
\[
I_w = \gamma h_{\mathbf{w}}  X_w||\mathbf{w}||^{-\alpha}
\]
where $h_w$ denotes the small-scale fading component. This is elaborated on the following remark.
\begin{remark}
In case the interfering links undergo large-scale shadow fading modeled using a log-normal distribution, the conditional detection probability is modified as
\begin{align}
        P^{\rm Shadow}_{\rm d} &=  \mathbb{E}\left[\exp{\left(\frac{-\gamma}{\xi_0 R^{-2\alpha}}\right)} \left( \prod_{{\bf w} \in \phi^{\prime}} \frac{1}{1+\frac{\gamma X_{\bf w} d_{\bf w}^{-\alpha}}{R^{-2\alpha}}} \right)\right], \nonumber \\
        &=\exp{\left(\frac{-\gamma}{\xi_0 R^{-2\alpha}}\right)} \left( \prod_{{\bf w} \in \phi^{\prime}} \mathbb{E}\left[ \frac{1}{1+\frac{\gamma X d_{\bf w}^{-\alpha}}{R^{-2\alpha}}}\right] \right)\label{eq:cond_succ_ln}
\end{align}
where the expectation is with respect to the log-normal random variable $X$. Since no closed-form exists for the log-normal case, this expectation can be evaluated numerically (e.g., via Gauss-Hermite quadrature~\cite{liu1994note}). The rest of the analysis, including the run-length distribution remains valid with this modified interference term.
\end{remark}
}

We characterize the conditional probability of successful target tracking based on the above result.
\begin{theorem}
Leveraging de Moivre's solution~\cite[Section 22.6]{hald2005history}, the conditional \ac{CCDF} ${F}_{L}(\nu)=\mathbb{P}\left(L \geq \nu \mid \Phi\right)$ of $L$ as a function of $P_{\rm d}$ is
    \begin{multline}\label{eq:CCDF_Restless}
     {F}_{L}(\nu)  = \sum_{l = 1}^{\lfloor \frac{T+1}{\nu+1} \rfloor} (-1)^{l + 1} \left(P_{\rm d} + \frac{T - l\nu +1}{l} (1 - P_{\rm d})\right)  \\ \times\binom{T- l\nu}{l - 1}P^{l\nu}_{\rm d} (1 - P_{\rm d})^{l - 1}. 
    \end{multline}
    \label{theo:demoivre}
\end{theorem}
\begin{IEEEproof}
    Please see Appendix~\ref{app:demoivre}.
\end{IEEEproof}

The above result establishes the probability of successful target tracking for a given realization $\phi$ of $\phi$ and a given access states for the active radars. Next, we look at the target tracking performance averaged over the network realization using the moments of conditional success probability.

\begin{lemma}\label{lem:moments}
Given that the ego radar is active in a block, the moments of conditional success probability $P_{\rm d}$ of the ego-radar at a given time $t$ within the block is given by
        \begin{align}
            &\zeta(l) = \mathbb{E}\left[P_{\rm d}^l\right] \nonumber \\
            &= \exp{\left(\frac{-l\gamma}{\xi_0 R^{-2\alpha}}\right)}\exp\left(- \lambda \delta \int_0^{\infty} 1 - \mathcal{T}_z^l{\rm d}z\right) \cdot \nonumber\\
            &\exp\left(-2\lambda_L\int_0^{2\pi}\int_{0}^\infty 1 - \exp\left( -\delta\lambda\int_{a_i}^{b_i} 1 - \mathcal{T}^l_{z,x} {\rm d}z \right){\rm d}x {\rm d}\theta\right),
        \end{align}
        where $\mathcal{T}_z = \left(\frac{R^{-2\alpha}}{R^{-2\alpha} + \gamma z^{-\alpha}}\right)$, $\mathcal{T}_{z,x} = \frac{R^{-2\alpha}}{R^{-2\alpha}+\gamma w(x,z,\theta))^{-\alpha}}$ and $w^2(x,z,\theta) = x^2 + z^2 + 2 x z \cos\theta$.
\end{lemma}
\begin{IEEEproof}
Please see Appendix~\ref{app:momentscsp}.
\end{IEEEproof}
The above result immediately leads to some interesting insights about the expected number of bursts that exceed the desired length and the expected burst length averaged across all network instances. 
\begin{theorem}
    The expected number of successful tracking events across all network instances is $(T - \nu +1)\zeta(\nu)$, and the expected tracking length is $\mathbb{E}[L] = \sum_{l = 1}^{T} \zeta(l) - T\zeta(T+1)$. 
    \label{prop:burstproperty}
\end{theorem}
\begin{IEEEproof}
Please see Appendix~\ref{app:burst}.
\end{IEEEproof}

\begin{corollary}
    For $T = \infty$ and the same arguments as that of the proof of \Cref{prop:burstproperty}, the expected tracking length is the sum of the moments $\sum_{l = 1}^{\infty} \zeta(l)$.
\end{corollary}

Further, we define the \ac{MD} of the burst length $L$ as the distribution of ${F}_{L}(\nu)$, i.e.,
    \begin{equation*}
        \cl{M}_{B}(\nu, \beta) = \mathbb{P}\left({F}_{L}(\nu) \geq \beta\right), 
    \end{equation*}
where ${F}_{L}(\nu) =\mathbb{P}\left(L \geq \nu \mid \Phi\right)$ is the conditional \ac{CCDF} of $L$ as defined in \eqref{eq:CCDF_Restless}. In other words, $\cl{M}(\nu, \beta)$ represents the fraction of radars in the network that experience a successful successive detection of at least $\nu$ slots in a sequence of $T$ slots in at least $\beta$ fraction of the network realization. Unlike the \ac{MD} of the SINR in wireless networks, we observe that the \ac{MD} of the tracking length is challenging to derive even indirectly via its moments. However, if all the radars have the same desired tracking length $\nu$, the first moment of the distribution is given by $\mathbb{E}\left[{F}_{L}(\nu)\right]$ in \Cref{theo:cond_ccdf_burst} below.

\begin{theorem}
Given that the ego radar transmits, the first moment of the conditional \ac{CCDF} of the tracking length, or the tracking probability is
    \begin{multline}\label{eq:blockprob}
        P_{\rm t} = \mathbb{E}\left[{F}_{L}(\nu) \right] = \sum_{l = 1}^{\lfloor \frac{T+1}{\nu+1} \rfloor} (-1)^{l + 1} \binom{T - l\nu}{l - 1}   \\
     \times \bigg( \sum_{{\ell} = 0}^{l - 1} \binom{l-1}{{\ell}} (-1)^{\ell} \zeta(l\nu + 1 +{\ell}) \\ +  \frac{T - l\nu +1}{l} \sum_{{\ell} = 0}^l \binom{l}{{\ell}} (-1)^{\ell} \zeta(l\nu +{\ell}) \bigg).
    \end{multline}
\label{theo:cond_ccdf_burst}
\end{theorem}
\begin{IEEEproof}
From \eqref{eq:CCDF_Restless}, the expected value of ${F}_{L}(\nu) $ is
\begin{multline*}
    \mathbb{E}\left[{F}_{L}(\nu) \right] = \sum_{l = 1}^{\lfloor \frac{T+1}{\nu+1} \rfloor} (-1)^{l + 1} \binom{T - l\nu}{l- 1} \\
     \times\mathbb{E}\left[P_{\rm d}^{l\nu+1} (1 - P_{\rm d})^{l-1} 
     +\frac{T - l\nu +1}{l} P_{\rm d}^{l\nu} (1 - P_{\rm d})^{l}\right].
\end{multline*}
Using the Binomial expansion $(1 - P_{\rm d})^{l} = \sum_{\ell = 0}^{l} \binom{l}{\ell} (-1)^{\ell} P_{\rm d}^{\ell}$ and some trivial simplification, we arrive at the result.
\end{IEEEproof}
{It is easy to see that the expression contains a finite number of terms. For example, for a block length of $T = 19$ slots and $\nu =9$ slots, $\lfloor \frac{T+1}{\nu+1} \rfloor = 2$. Accordingly, the number of total terms in the summation is 8. Recall that the conditional detection probability $P_{\rm d}$ is upper bounded by 1 (since it is a probability). Thus, $P_{\rm d}^l \leq 1$ and consequently, due to monotonicity of expectations, we have $\mathbb{E}[P^l_{\rm d}] =  \zeta(k) \leq 1$. We complete the argument by stating that since the expression of Theorem~\ref{theo:cond_ccdf_burst} consists of a finite number of bounded terms, the expression for $P_{\rm t}$ converges trivially.
}

\section{\ac{QoS}-Aware Optimal Frame Design}
\label{sec:OFD}
In this section, we use the above framework for the design of the transmit frames and the selection of optimal access probability from the perspective of specific automotive use-cases. Let us define $T_{\rm L}$ as the latency requirement specified by the use-case. As an example, low-latency applications such as lane change assistance and automatic emergency braking (AEB) may employ a $T_{\rm L}$ of the order of 1 ms (note that this not the end-to-end latency, but only the latency over the radio link). On the contrary, use-cases that have less stringent latency requirements such as blind spot detection, park assist etc have $T_{\rm L}$ of the order of 10 ms. Some indicative latency requirements are outlined in Table~\ref{tab:tab1}. Furthermore, we introduce a reliability measure based on the parameter $\nu$. in particular, a larger value of $\nu$ corresponds to a higher reliability requirement from the service. For the tracking service to be rendered successful, let us assume that within the latency deadline, the use-case necessitates at least one successful tracking event, i.e., the ego radar experiences at least one burst of $\nu$ consecutive successes within a deadline of $T_{\rm L}$.

As per the radar \ac{MAC} introduced in this work, let the duration $T_{\rm L}$ be divided into $N$ access blocks, wherein, each radar competes for access in each block. Recall that each block is further divided into $T$ slots, each corresponds to one pulse width $W$ of the radar. Naturally, the choice of $N$ (and consequently, $T$) depends on the required $\nu$ of tracking, which in turn, depends on the use-case. In particular, we have
\begin{align}
    N = \Big\lfloor {\frac{T_{\rm L}}{TW}} \Big\rfloor,
\end{align}
where $T$ must be necessarily larger than $\nu$. For a specific use case, the probability that the target is successfully tracked in at least one of the tracking attempts within the latency deadline is thus $(1 - (1 - \delta)^N)P_{\rm t}$. This naturally incorporates the spatial correlation of each network realization since for a fixed $\delta$, $\mathbb{E}[(1 - (1 - \delta)^N) \mathbb{P}(L \geq \nu | \Phi)] = (1 - (1 - \delta)^N)\mathbb{E}[F_L(\nu)]$.

A natural trade off arises - fewer blocks results in a lower access chance for individual radars for activation during $T_{\rm L}$, i.e., a lower value of $(1 - (1 - \delta)^N)$. However, it results in a larger $T$ in each block which enables a higher probability that the tracking length of an active radar exceeds $\nu$, i.e., a larger $P_{\rm t}$. Two extreme cases are of special interest - one where the entire latency period is allotted to all the radars for access, i.e., we set $TW = T_{\rm L}$. The other case is to set $T = \nu$, i.e., the block length is exactly equal to the reliability requirement.

An optimal frame design along with an optimal access policy needs cognitive information about the number of competing radars. It is mathematically formulated as follows.
\begin{align}
    \max_{\delta, N} \quad  & (1 - (1 - \delta)^N)P_{\rm t} \nonumber \\
    \text{st.} \quad  &P_{\rm t} \text{ as in } \eqref{eq:blockprob}\nonumber \\
    &N = \bigg\lfloor\frac{T_{\rm L}}{TW}\bigg\rfloor \leq \bigg\lfloor\frac{T_{\rm L}}{\nu W}\bigg\rfloor \nonumber \\
    & 0 \leq \delta \leq 1 \nonumber    
\end{align}
The above formulation is a mixed integer non-linear program, where the parameters $\lambda$ and $\lambda_L$ may not be known to the ego-radar. This necessitates the development of data-driven algorithms for optimizing the tracking performance, which is currently out of scope of this paper where our focus is mainly on the statistical characterization. Nevertheless, we make a couple of critical remarks.
\begin{remark}
     The case $\delta = 1$ corresponds to full interference (no \ac{MAC} protocol). In this case, regardless of the value of $\nu$, the optimal solution is to select $N = 1$ and accordingly $TW = T_{\rm L}$.
     \label{remark:full_interference}
\end{remark}
Indeed, for the full interference scenario, any other choice of $N$ simple reduces the block length without improving the tracking probability. The case $\nu = 1$ corresponds to maximizing the detection probability, i.e., at least one successful detection in the block. We will see in the next section that the optimal $\delta$ decreases with an increase in $\lambda$ and $\lambda_L$ due to an increase in interference. Hence, for a given block size, block length, and $\nu$, the optimal $\delta$ can be read out from an appropriate look-up table.
\begin{remark}
    The optimization over $N$ is simple since the search space is finite. Starting from $N = 1$, each step increase in $N$ reduces the block length by half. Thus, the search space for $N$ is $\left[1,\log_2\left(T_L/W\right)\right]$.
\end{remark}
In case of a finite set of access probabilities, as is the case in most practical deployments, the optimal pair of $N$ and $\delta$ can thus be stored for given $\lambda$ and $\lambda_L$.

\section{Numerical Results and Discussion}
In this section we discuss the salient features of the network with the help of some numerical results. The radar parameters used for the numerical results are taken from~\cite{series2014systems}. Specifically, $P = 10$ dBm, $\bar{\sigma} = 30$ dBsm, $\alpha = 2$, $G_{\rm t} = 10$ dBi, $f_c = 76.5$ GHz, and $N_0 = -174$ dBm/Hz. Unless  otherwise stated, we assume $R=15$ m. The street and vehicle densities are from \cite{shah2024modeling} where the authors have reported real-world data on the same.
\label{sec:NRD}
\subsection{Detection vs Tracking Performance}
\begin{figure}
    \centering
    \includegraphics[width=0.75\linewidth]{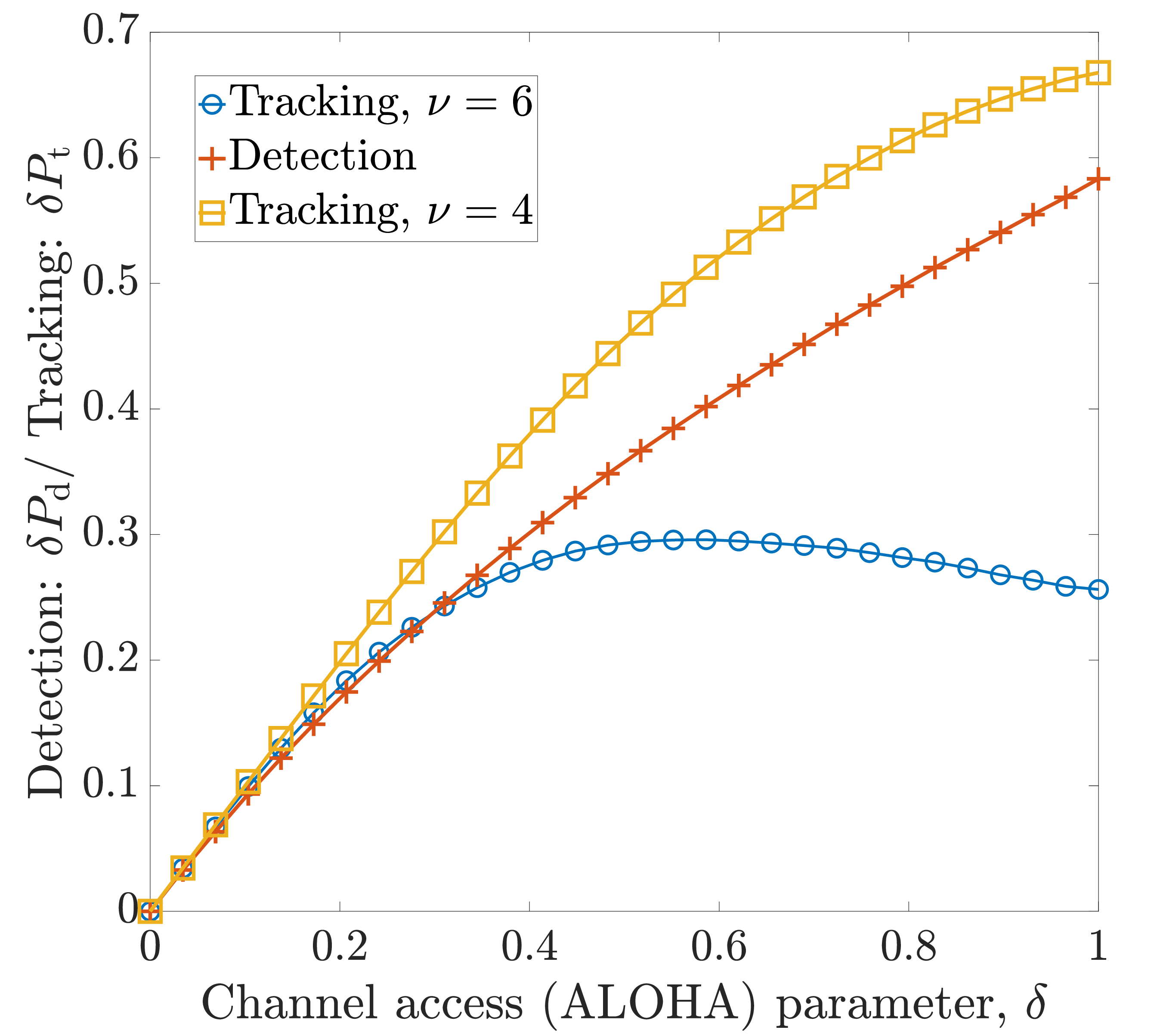}
    \caption{Comparison of detection and tracking performance with respect to the ALOHA parameter. Here $\lambda = 5e-2$ m$^{-1}$, $T = 20$ slots, $\nu = 6$, and $\lambda_L = 5e-4$ m$^{-1}$.}
    \label{fig:Det_vs_Track}
\end{figure}

First, we discuss the key insights that our analysis provides as compared to the existing works on stochastic geometry based automotive radar analysis. In Fig.~\ref{fig:Det_vs_Track} we plot the radar detection probability in one slot multiplied with the channel access probability, i.e., $\delta P_{\rm d}$ and compare it with the tracking probability in one block multiplied with the block access probability, i.e., $\delta P_{\rm t}$. Based on the reliability requirement, $\nu$, the tracking probability may vary drastically. A low $\nu$ enables a higher tolerance of interference. In particular, we see that for $\nu = 4$, the framework recommends an optimal $\delta = 1$, as even in the full interference regime, the event $L \geq \nu$ is achieved with the highest probability. On the contrary, a more stringent reliability requirement, e.g., $\nu = 6$ necessarily requires access control - here the optimal $\delta$ is near 0.5. In other words, even though half of the radars do not access the channel on an average, the resultant lower interference results in an improved tracking probability. We further note that $\nu = 4$ has an improved tracking probability as compared to the detection probability. For explaining this, is worth to recall that as per the block MAC protocol introduced in this paper, once a radar is active in a block, it transmits in all the slots of the block. This is contrary to the detection event where the radar competes for access in all the slots of the block. Hence, for a lower value of $\nu$, a successful channel access in a block may be sufficient for a successful tracking event, especially in the low $\delta$ (low interference) regime.

\begin{figure}
    \centering
    \includegraphics[width=0.75\linewidth]{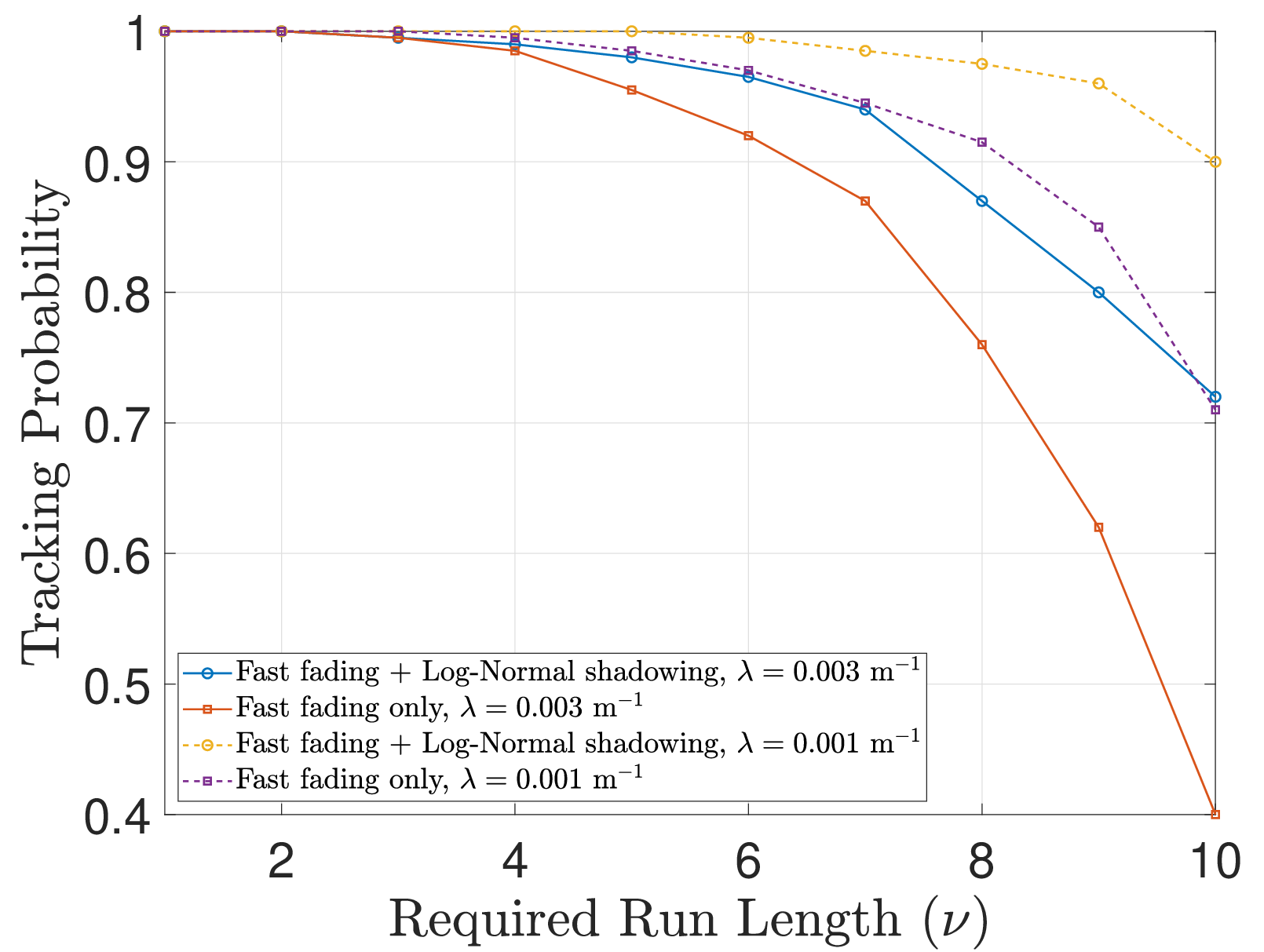}
    \caption{Impact of shadowing on the trends of the tracking probability with respect to $\nu$.}
    \label{fig:LNvsRay}
\end{figure}

\subsection{Impact of Shadowing and System Parameters}
{\color{blue}Fig.~\ref{fig:LNvsRay} shows that log-normal shadow fading deteriorates the interference signals thereby improving the tracking probability. In general, the ego-radar intends to track the vehicle that is directly in front or in the rear of it. Accordingly, it is a reasonable assumption that the target vehicle is in line-of-sight. Thus, shadowing improves the tracking performance without deteriorating the useful reflected signal.}

\begin{figure}
    \centering
    \includegraphics[width=0.75\linewidth]{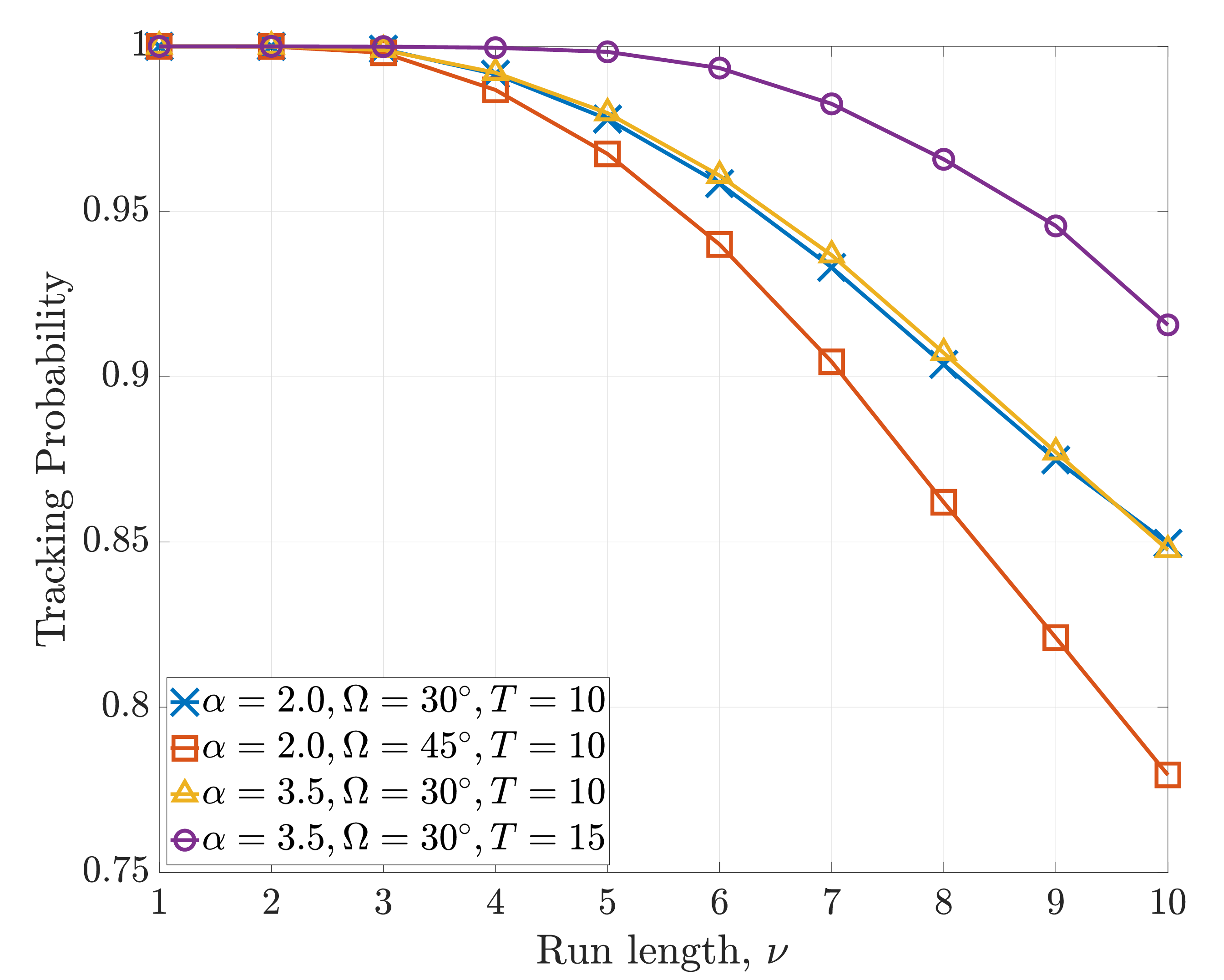}
    \caption{Impact of the system parameters on the tracking probability with respect to $\nu$. Here $R = 10$ m.}
    \label{fig:Param}
\end{figure}

{\color{blue}To illustrate the effect of key parameters on the tracking probability, Fig.~\ref{fig:Param} plots the probability of maintaining a run of length $\nu$ under different values of the path-loss exponent $\alpha$, beamwidth $\Omega$, and frame length $T$. We observe that increasing the beamwidth (e.g., from $\Omega=30^\circ$ to $45^\circ$) significantly reduces the tracking probability, as the wider beam admits more interferers. Conversely, increasing the path-loss exponent (from $\alpha=2.0$ to $\alpha=3.5$) improves the tracking probability by attenuating the interference more strongly - the impact on the two-way reflect signal is lower due to the low target distance. Finally, increasing the frame length $T$ enhances the probability of sustaining longer runs, thereby yielding better tracking performance. These trends confirm the sensitivity of radar tracking performance to physical-layer and system design parameters, and highlight the importance of careful parameter tuning.}

\subsection{Impact of Street and Vehicle Density}

\begin{figure}
    \centering
    \includegraphics[width=0.75\linewidth]{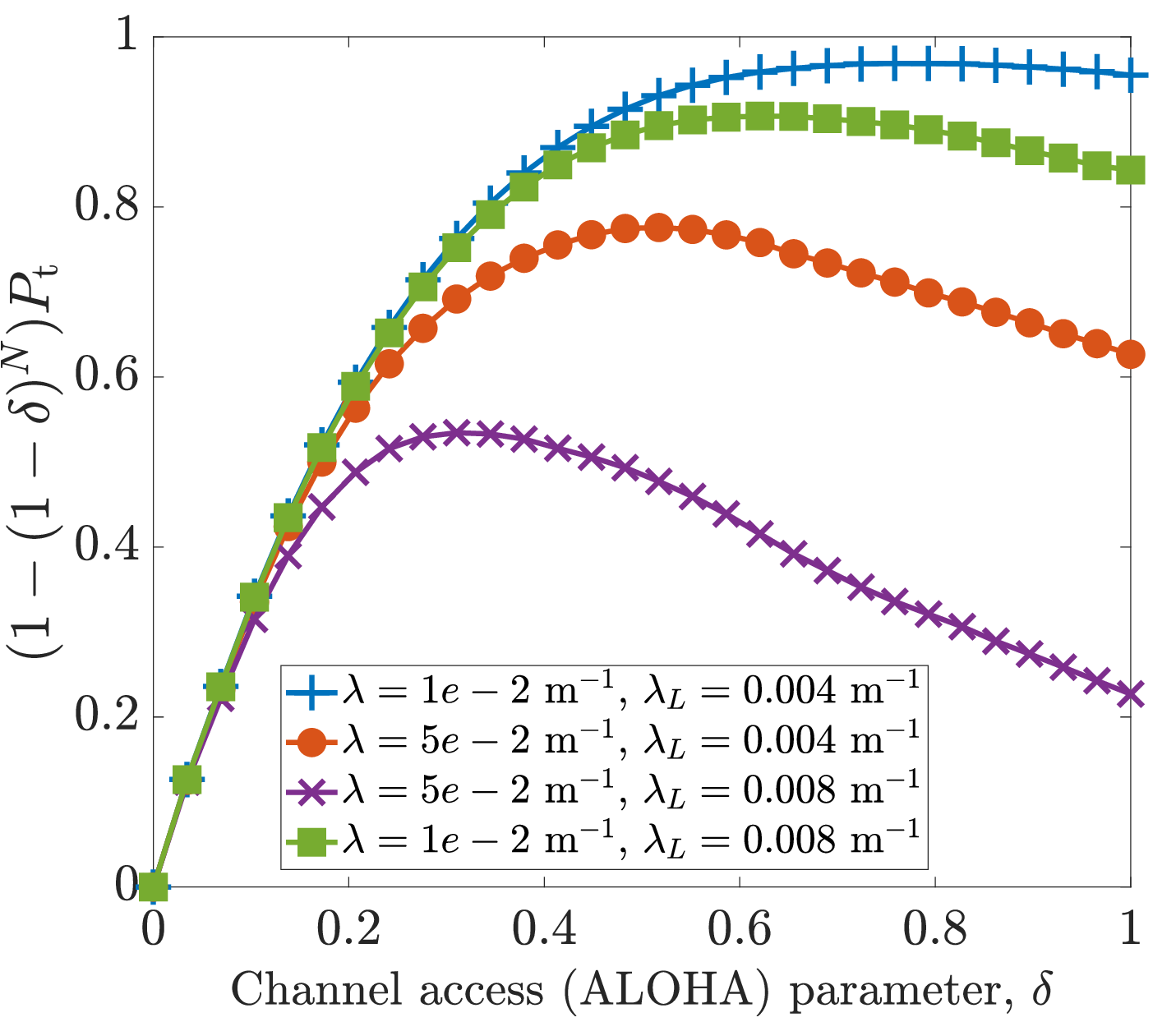}
    \caption{Impact of the street and vehicle density on the tracking performance. Here $T = 20$ slots, $K = 4$, and $\nu = 6$.}
    \label{fig:With_Vehicle}
\end{figure}

Fig.~\ref{fig:With_Vehicle} shows that the street geometry and the vehicle density influences not only the optimal channel access probability but also dictates the trends of the tracking performance $(1 - (1 -\delta)^N P_{\rm t})$. For dense streets, e.g., in the city center, an increase in $\delta$ leads to a higher increase in the interference as compared to scenarios with fewer streets. Accordingly, dense street scenarios necessitate stringent access control, facilitated by a lower value of $\delta$. A higher vehicle density reduces the tracking probability of the ego radar. This also necessitates the employment of a carefully designed MAC policy.
\begin{figure}
    \centering
    \includegraphics[width=0.75\linewidth]{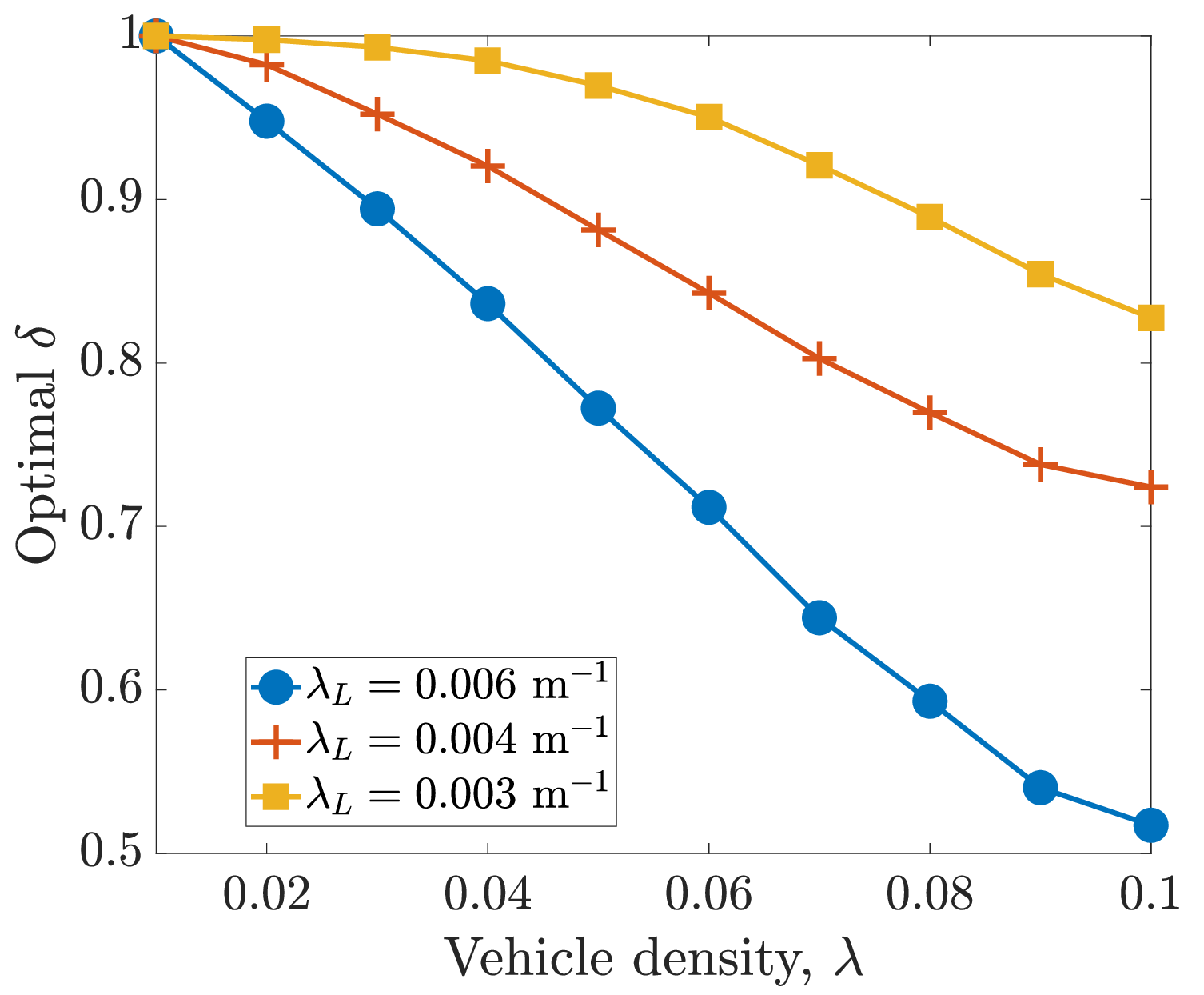}
    \caption{Optimal $\delta$ for different vehicle and street density. Here we fix $T = 25$ and $N = 4$.}
    \label{fig:optimal}
\end{figure}
This is illustrated in Fig.~\ref{fig:optimal} where we see that the optimal $\delta$ decreases with an increase in either $\lambda$ or $\lambda_L$. Although the optimal $\delta$ for low $\lambda$ is 1 for both $\lambda_L = 0.004$ m$^{-1}$ and $\lambda_L = 0.006$ m$^{-1}$, the optimal $\delta$ decreases more rapidly for a higher value of $\lambda_L$.
\begin{figure}
    \centering
    \includegraphics[width=0.75\linewidth]{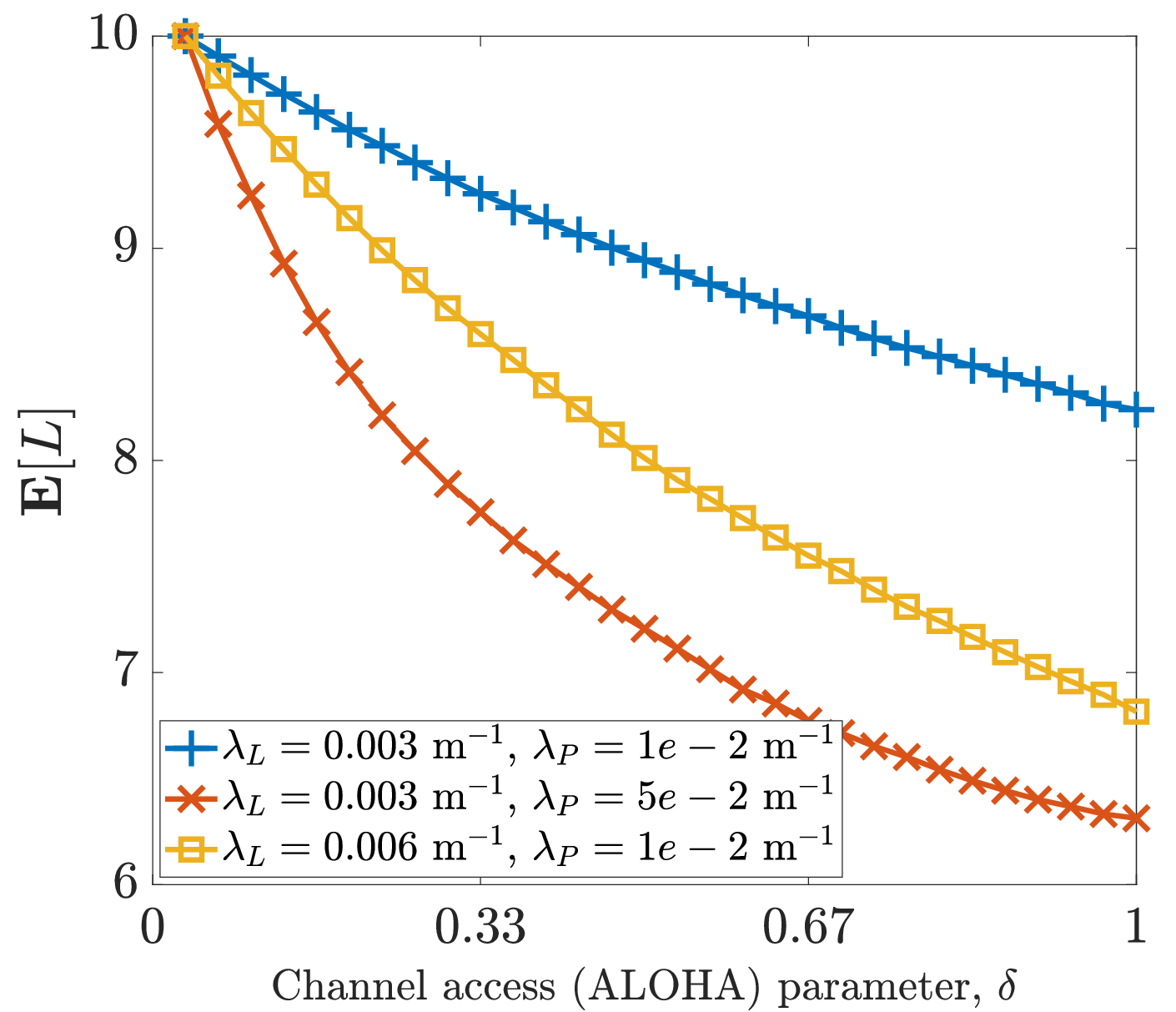}
    \caption{Expected number of bursts with a length $\geq \nu = 8$ for $T = 20$ for different $\lambda_P$ and $\lambda_L$.}
    \label{fig:QoS4}
\end{figure}
Furthermore, Fig.~\ref{fig:QoS4} shows the trends in the conditional expectation of the number of successful bursts of $8$ or more successive successful detection events in a block of length $20$. The condition is on the event that the ego-radar accesses the channel in the block under consideration. Note that the maximum value of the conditional expectation in this case is $10$. Higher the ALOHA parameter, lower are the number of successful bursts experienced by the ego-radar. The number of successful bursts required depend on the use-case QoS requirement and must be carefully studied. This is more crucial since the analysis in Fig.~\ref{fig:QoS4} considers overlapping bursts, which may or may not be suited for a given service. This is further discussed below.

\begin{table*}[h!]
\centering
\begin{tabular}{| m{5cm} | m{2cm} | m{2.8cm} | m{2cm}|}
\hline
{Use Case} & {Latency} $T_{\rm L}$ & {Reliability,} $\nu$ & $\delta^*$ \\
\hline
Adaptive Cruise Control (ACC) & $\leq$ 100 ms & 10 &1 \\
\hline
Automatic Emergency Braking (AEB) & $\leq$ 50 ms & 15 &0.8\\
\hline
Blind Spot Detection (BSD) & $\leq$ 200 ms & 5 &1\\
\hline
Lane Change Assistance & $\leq$ 100 ms & 15& 0.4\\
\hline
Pedestrian Detection & $\leq$ 50 ms & 15 &0.8\\
\hline
Park Assist & $\leq$ 100 ms & 1 &1\\
\hline
Intersection Management & $\leq$ 50 ms & 15 &0.8\\
\hline
\end{tabular}
\caption{Use cases and corresponding latency and reliability requirement derived from~\cite{boban2018connected} and \cite{3gpp2019service}. However, note that the latency requirement mentioned here is not end-to-end latency, but the latency over the radio link. The figures are indicative and real values are system dependent. The optimal channel access probability is calculated against each use-case. We assume $\lambda = 5e-2$ m$^{-1}$ and a beamwidth of 10 degrees.}
\label{tab:tab1}
\end{table*}

\subsection{Impact of Use-Case QoS Requirements}
\begin{figure}
    \centering
    \includegraphics[width=0.75\linewidth]{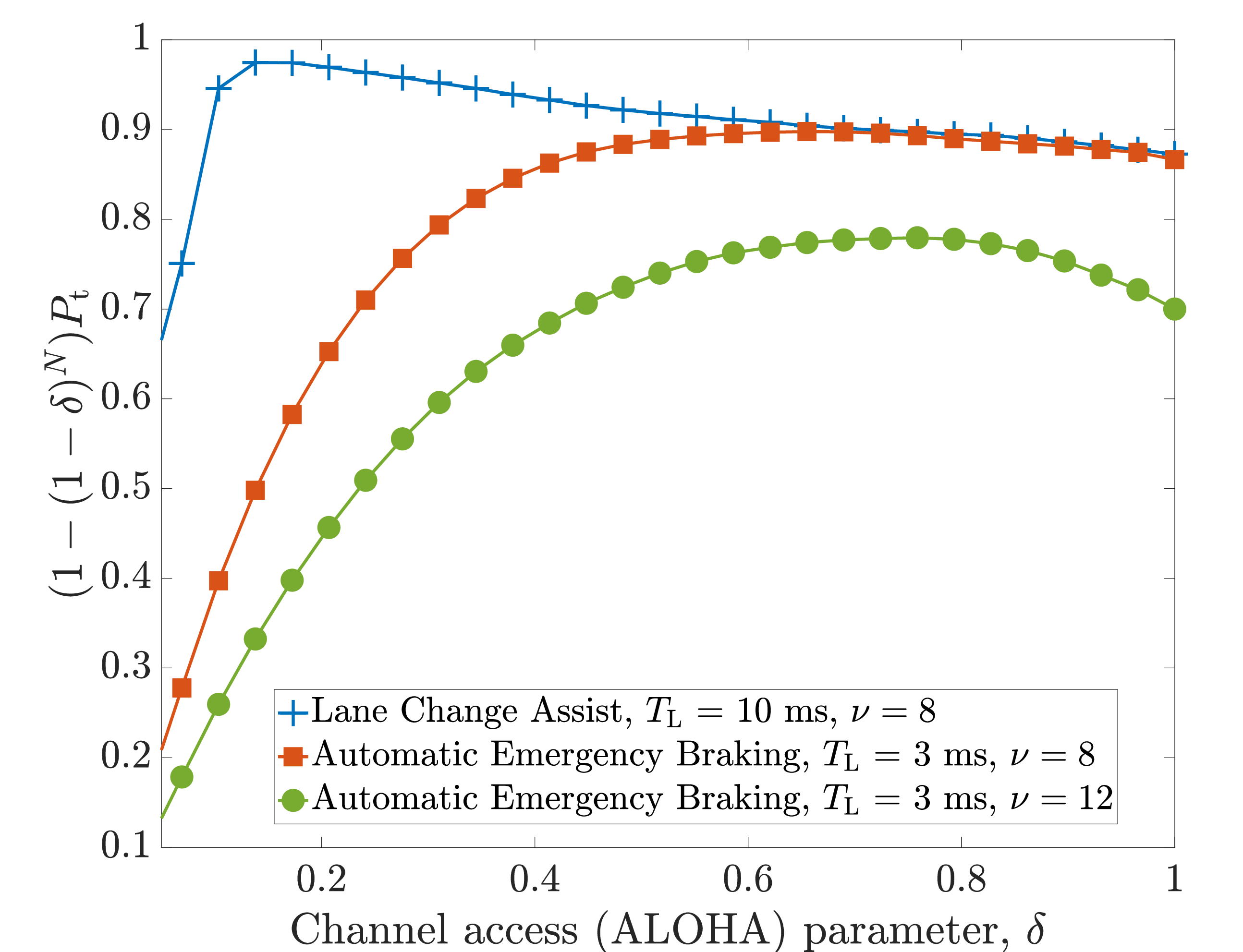}
    \caption{Impact of the use-case QoS requirements on the tracking performance. Here we fix $T = 20$ slots.}
    \label{fig:QoS1}
\end{figure}

Fig.~\ref{fig:QoS1} shows the impact of the latency deadline $T_{\rm L}$ and the reliability requirement modeled as the number of consecutive slots of detection $\nu$. Specifically, we study two services - lane change assist~\cite{alonso2008lane} with a typical latency requirement of the order of 10 ms, and automatic emergency braking~\cite{sidorenko2021safety} with a typical latency of 1-3 ms. Note that these latency requirements from the radio link and not end-to-end latency. Furthermore, let us specify two reliability requirements: $\nu = 8$ and $\nu = 12$. For the results of this figure, we fix the number of slots per block as $T = 20$. We assume a pulse width of $200$ $\mu$s. This results in $K = 50$ blocks for the lane change assist use-case and $k = 15$ blocks for the automatic emergency braking use-case. Recall that for the service to be rendered successful, the ego radar needs at least one successful tracking event. Due the larger number of blocks to compete in for access, the success probability is higher for the lane change assist use-case as compared to the automatic emergency braking use-case. Interestingly, for higher reliability requirement ($\nu = 12$), not only the success probability is lower, but also the optimal $\delta$ is higher. In Table~\ref{tab:tab1} we show example use-cases with corresponding latency and reliability requirements. Accordingly, the optimal channel access probability is demonstrated. The optimization of $\delta$ is not straightforward due to the two competing phenomenon of interference and channel access. in this regard, we discuss the considerations for the optimal frame design next.

\subsection{Optimal Frame Design}

\begin{figure}
    \centering
    \includegraphics[width=0.75\linewidth]{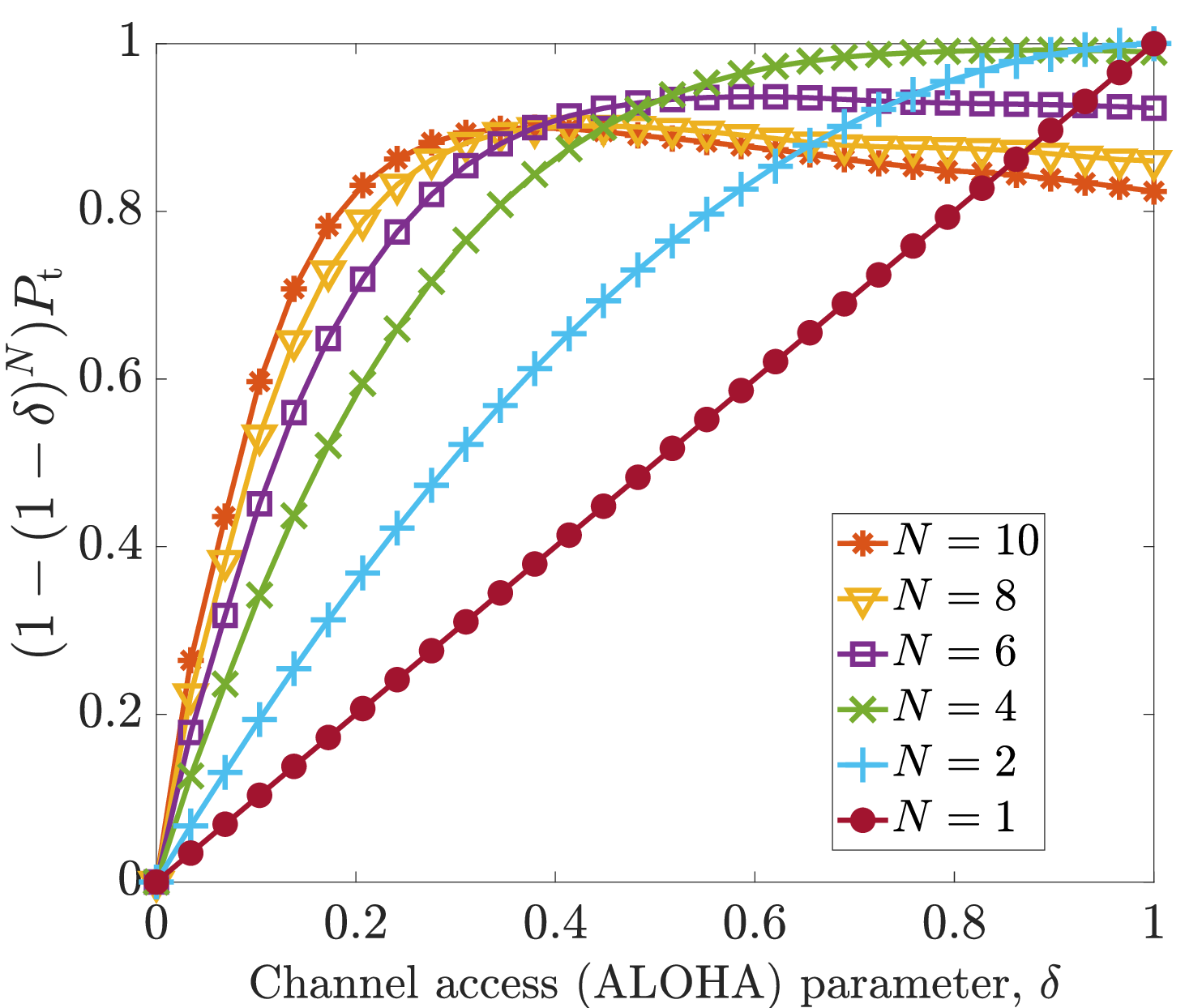}
    \caption{Impact of the frame design policy on the tracking performance. Here we use real data fitting of New Delhi city to extract the PLCP parameter $\lambda_L = 0.004$ m$^{-1}$ (see~\cite{shah2024modeling} for details) and use a vehicle density of $\lambda = 5e-2$ m$^{-1}$ and $\Omega = 10$ degrees.}
    \label{fig:QoS2}
\end{figure}
As discussed in the previous section, a larger value of $N$ results in smaller block lengths. This facilitates a larger number of attempts for the ego-radar to access transmission, but puts a more stringent demand on the tracking event. In this section, we investigate the impact of the beamwidth on optimal frame design. For a beamwidth of 10 degrees, Fig.~\ref{fig:QoS2} shows that for lower values of $\delta$, a higher value of $N$ results in an improved tracking performance. Indeed, when fewer interferers are active (lower $\delta$), a limited number of access opportunities for the ego radar can result in a successful target tracking. However, in case of high interference (higher $\delta$), a larger number of access attempts are necessary. As noted in Remark~\ref{remark:full_interference}, for $\delta = 1$, higher the value of $N$, lower is the tracking success. For this particular case, we see that $N = 1$ and $\delta = 1$ is optimal (in terms of $(1 - (1 - \delta)^N)P_{\rm t}$.

\begin{figure}
    \centering
    \includegraphics[width=0.75\linewidth]{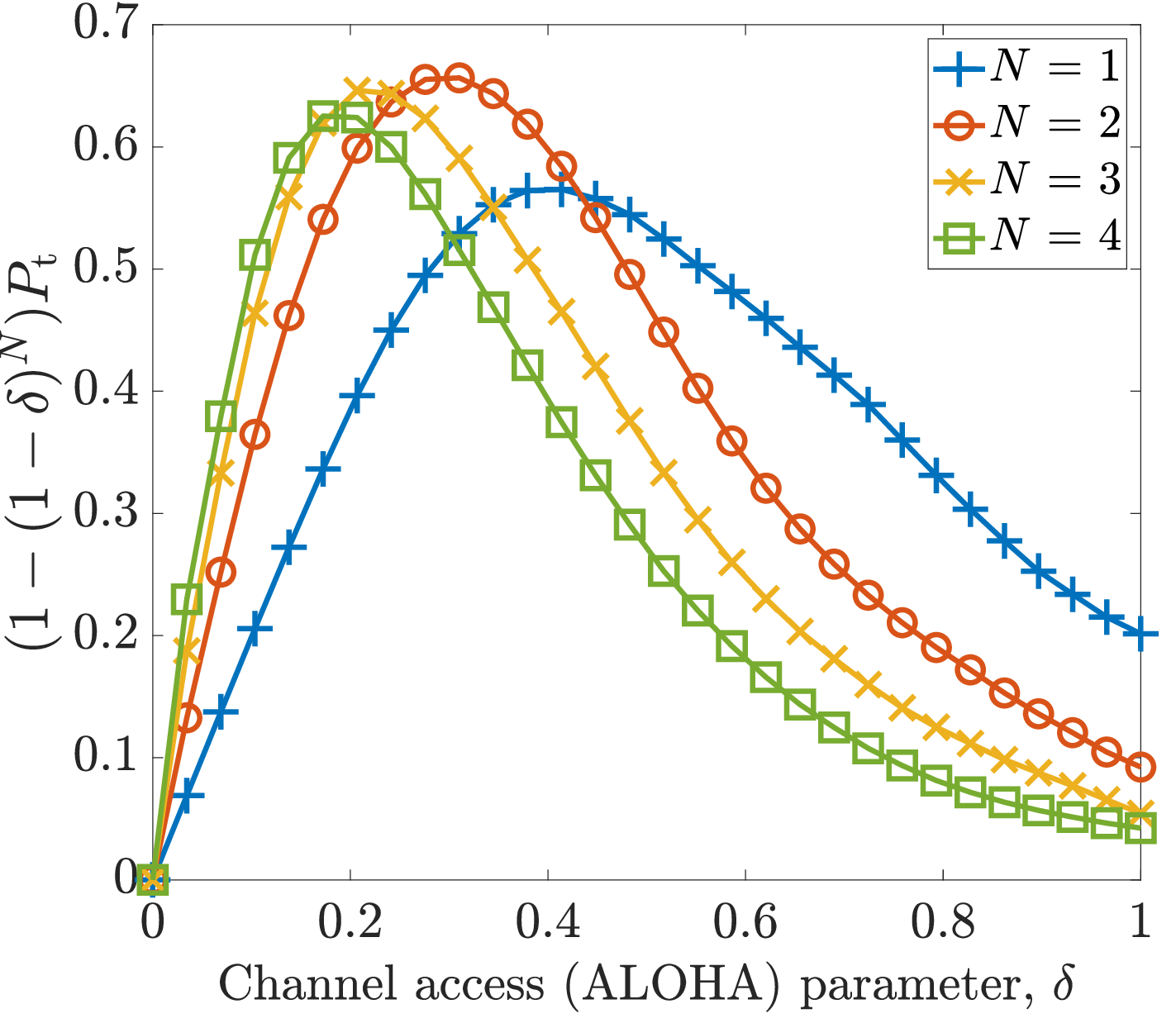}
    \caption{Impact of the frame design policy on the tracking performance for a wide beamwidth: $\Omega = 50$ degrees.}
    \label{fig:QoS3}
\end{figure}
In case of a higher beamwidth, e.g., $\Omega = 50$ degrees, Fig.~\ref{fig:QoS3} shows that $N = 1$ may not be optimal.  A larger $\Omega$ corresponds to a larger number of interfering radars appearing within the radar sector of the ego-radar, albeit with a lower radiated power. In this case, we see that $N = 2$ results in a higher tracking performance when $\delta$ is optimally chosen (0.3) in this case. Similar to the previous case, the optimal $\delta$ is lower for a larger value of $N$. This follows naturally since a higher $N$ results in a shorter blocks size, which necessitates a more stringent access control to limit the interference.

    \begin{figure}
    \centering
    \includegraphics[width=0.75\linewidth]{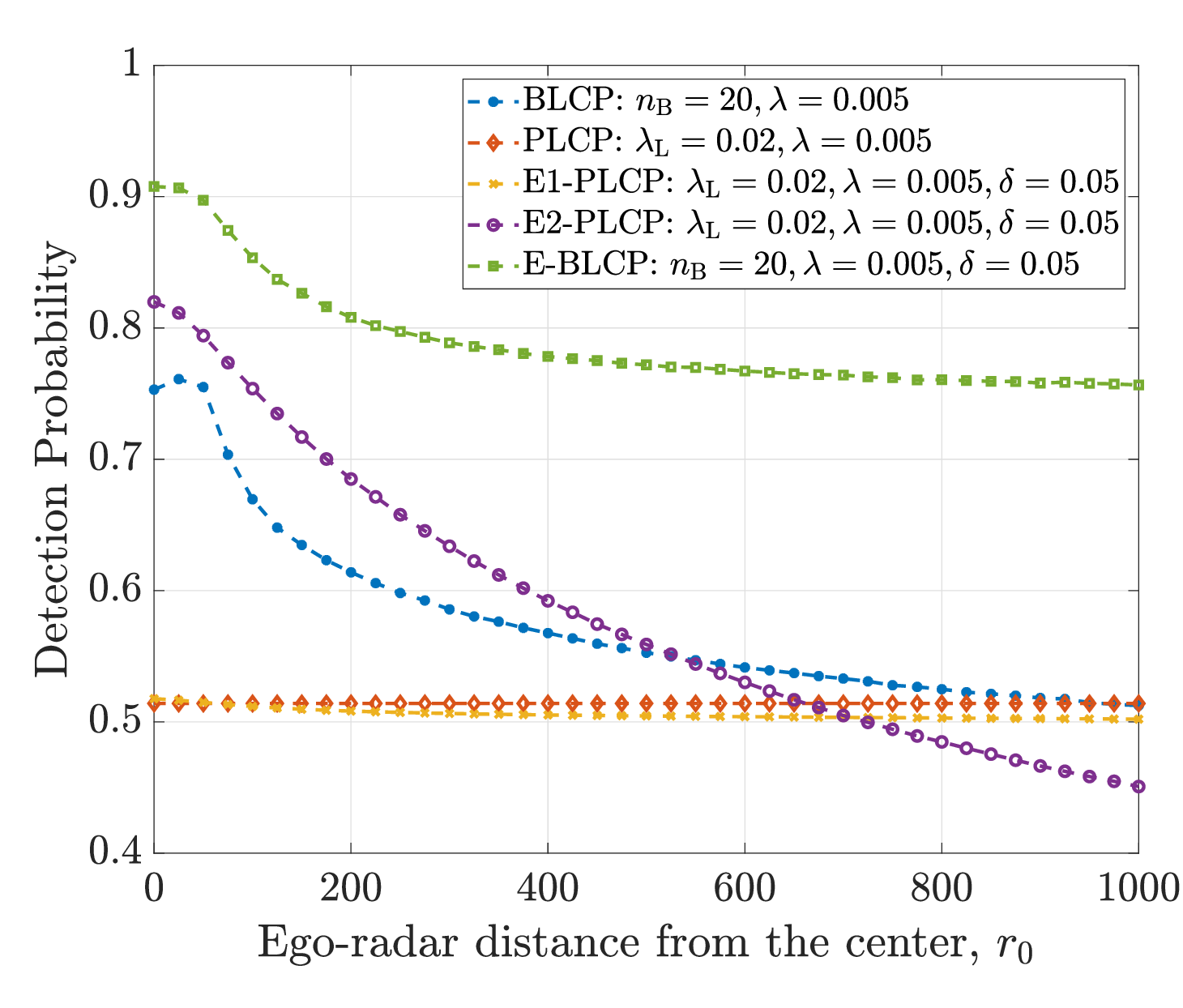}
    \caption{Trends of the detection probability for homogeneous PLCP vs inhomogeneous PLCP and BLCP with respect to the distance of the ego-radar from the center of the city.}
    \label{fig:plot}
\end{figure}

\begin{figure}
    \centering
    \subfloat[]
    {\includegraphics[width=0.45\linewidth]{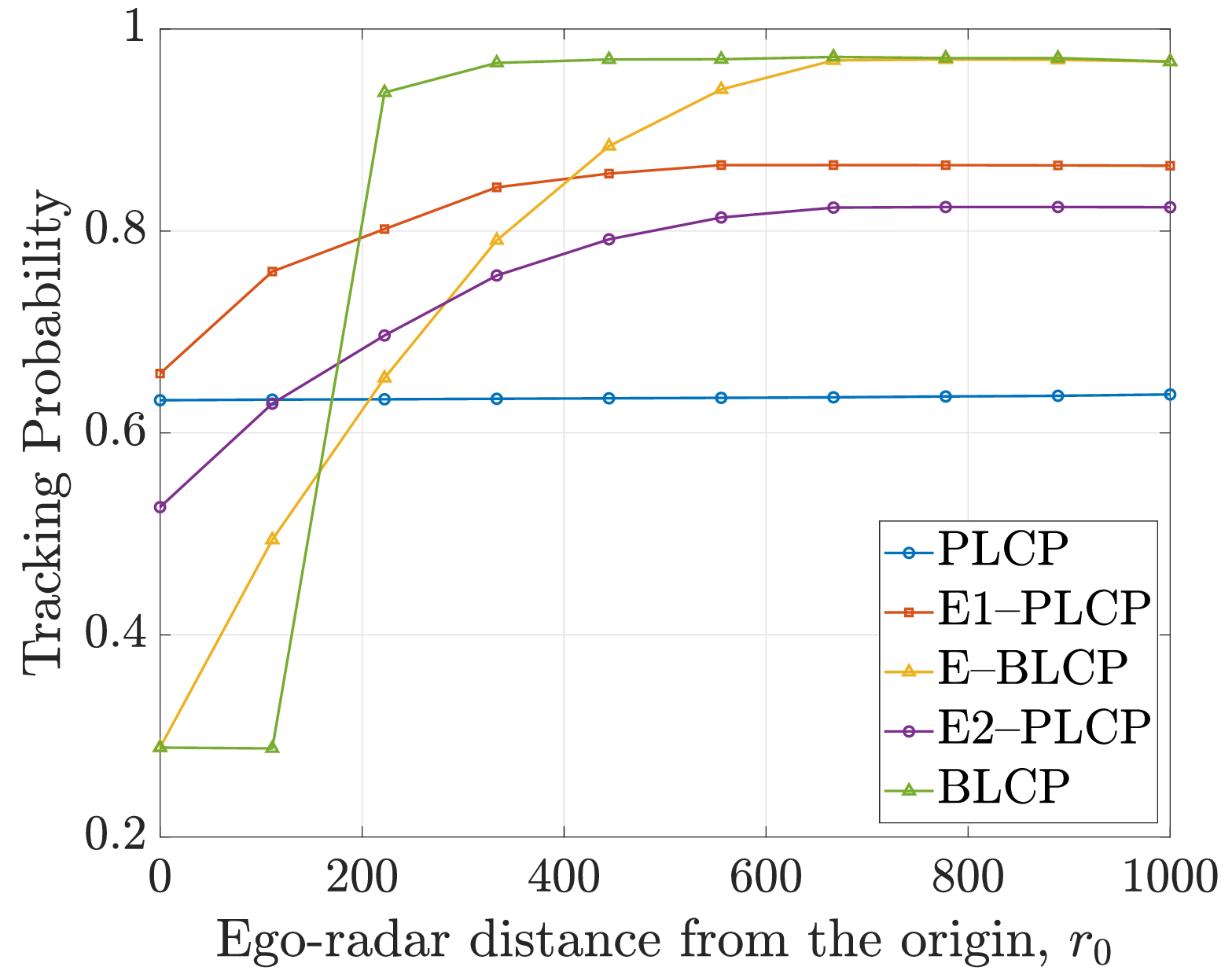}\label{fig:Inhom1}}
    \hfil
    \subfloat[]
    {\includegraphics[width=0.45\linewidth]{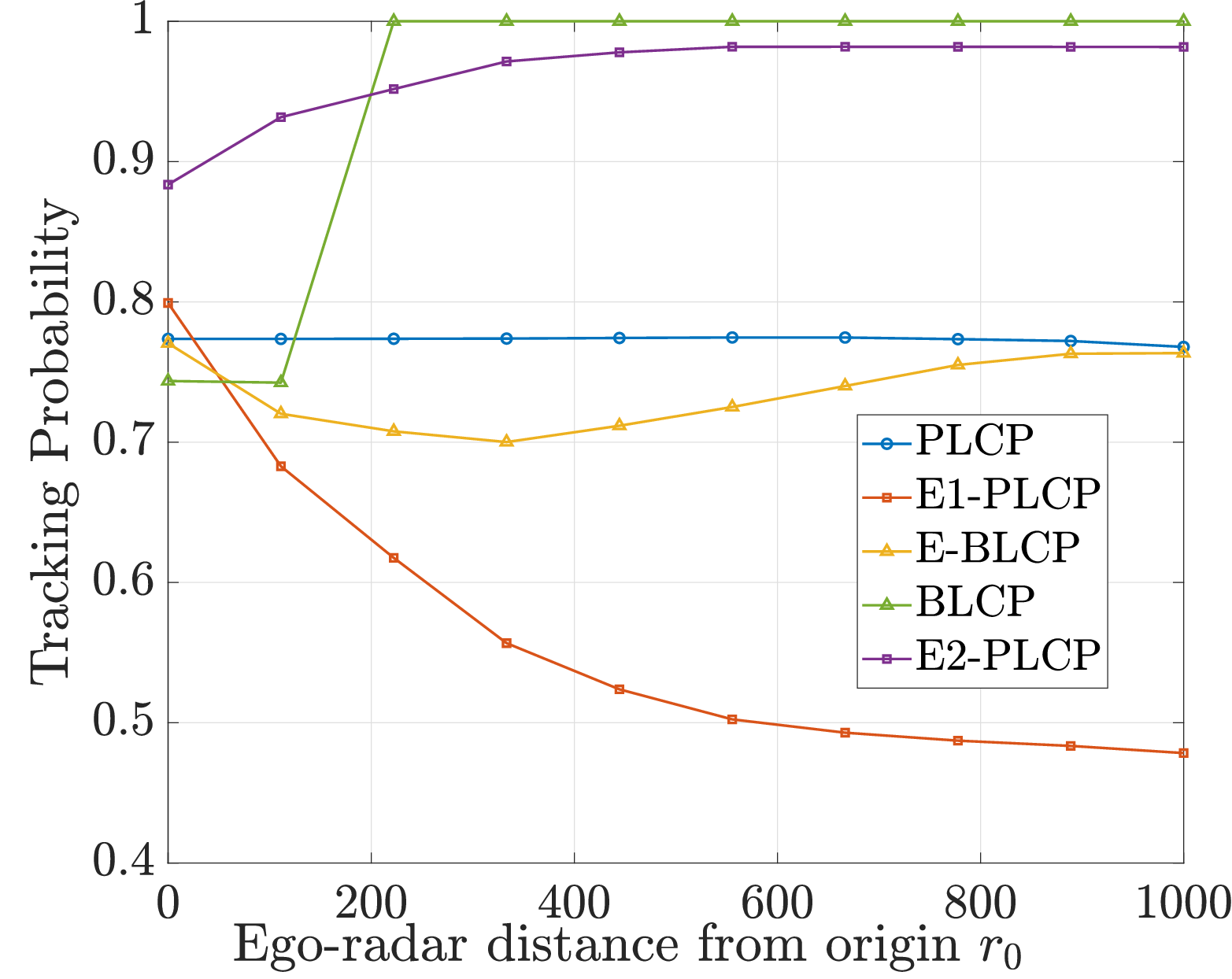}\label{fig:Inhom2}}
    \caption{Trends of the tracking probability for homogeneous PLCP vs inhomogeneous PLCP and BLCP with respect to the distance of the ego-radar from the center of the city for (a) fixed target distance and (b) stochastic target distance.}
    \label{fig:Inhom}
\end{figure}
\subsection{Impact of Non-Stationarity}
{\color{blue}Street systems are denser near the city center while becoming sparse in the outskirts. One such non-stationary aspect of street networks was recently taken up in \cite{shah2024binomial} where a new model called \ac{BLCP} was introduced which models inhomogeneous street networks across different cities. Remarkably, the model was validated using real data collected from maps in~\cite{11075590}. Unlike the PLCP model, in an inhomogeneous process, there does not exist a typical point from the perspective of which the performance can be characterized. On the contrary, the performance is bench-marked for a {\it test} point conditioned on its location (say $r_0$ from the origin). As suggested by the reviewer, we investigate the change in the trends of the tracking probability as the model becomes increasingly non-stationary. Additionally, considering that the vehicular nodes too become sparser as we move out of the city, we assume that the ego-vehicle intends to detect its nearest target. Consequently, we model the location of the target as the nearest point of the doubly stochastic Cox processes.

In Fig.~\ref{fig:plot}, we show the trends of the detection probability for $r_0$ for three new Cox process models, the newly introduced \ac{BLCP} model (e.g., see~\cite{shah2024binomial}) and contrast it with the homogeneous \ac{PLCP}. Among the contending models, the E1-\ac{PLCP} model is constructed by thinning the homogeneous \ac{PLCP} points based on their distance from the origin using an exponential decay function, $f(x) = \exp{(-\delta x)}$, where $x$ represents the point's distance from the origin and $\delta$ is the decay factor. Similarly, in the E2-\ac{PLCP} model, the intensity of the line process is varied as per an exponential-based intensity function, $\lambda(r) = \exp{(-\delta r)}$, where $r$ is the perpendicular distance of a \ac{PLCP} line from the origin. In the E-\ac{BLCP} model, an additional layer of randomness is added to the \ac{BLCP} process by generating non-homogeneous PPPs on each line with intensity function $\lambda(x) = \exp{(-\delta x)}$, where $x$ denotes the distance of a \ac{BLCP} point from the generating point of its respective line. This results in a more realistic distribution of points concentrated around the city center, enhancing the realism of the original \ac{BLCP} model. Fig.~\ref{fig:plot} shows that for \ac{BLCP} and E-\ac{BLCP}, the detection success probability initially rises with $r_0$, peaks, and then decreases. The point of minimum value of $\mathbb{E}\left[\frac{R}{d_1}\right]$, where $R$ is the target distance and $d_1$ is the closest interferer distance is the same as the value where success probability is maximum. It is worth recalling here that now both $d_1$ and $R$ are random quantities. In the case of homogeneous \ac{PLCP}, naturally, the probability of success is constant at all locations.

We extend this discussion to the tracking probability in Fig.~\ref{fig:Inhom}, where we plot the cases of fixed target distance (Fig.~\ref{fig:Inhom1}) and random target distance (Fig.~\ref{fig:Inhom2}). Here for the BLCP model, we have assumed the radius of the generating circle as 100 m. For the fixed target distance, the we see that the tracking probability increases as the ego-radar goes out of the city center and towards the outskirts. This is due to the decrease in interference as a result of less dense presence of the vehicles as $r_0$ increases. However, for the random target distance, as the vehicles become less dense, not only the interference decreases but the target also moves statistically farther from the ego-radar for the E1-PLCP and E-BLCP models, and hence we note a deterioration of the tracking probability. In a realistic deployment, the ego-radar can be equipped with a change-detection module (e.g., see~\cite{ghatak2021kolmogorov}) to sense the vehicle density and adapt its radar parameters accordingly.}

\subsection{Implementation Considerations}
{\color{blue}
\begin{table*}[h]
\centering
\caption{Execution time benchmarks for evaluating the analytical expressions.}
\begin{tabular}{|c|c|c|}
\hline
\textbf{Implementation} & \textbf{Average runtime per evaluation} & \textbf{Relative to 20~ms frame} \\
\hline
MATLAB (prototype)   & 0.40 ms & 2\% of frame \\
Python (NumPy/SciPy) & 0.30 ms & 1.5\% of frame \\
Optimized C (single-core) & 0.18 ms & 0.9\% of frame \\
Optimized C (multi-core, 4 threads) & 0.08 ms & 0.4\% of frame \\
\hline
\end{tabular}
\label{tab:comp}
\end{table*}
}
\textcolor{blue}{Real-time feasibility benchmarks:} {It is worth to highlight that the implementation of the protocol proposed in this work involves solving multiple integrals and summation with on-board processors in automotive systems, which can be challenging. However, with recent advancements of radar processing units appended with hardware-accelerated processing (e.g., see the discussion in ~\cite{engels2021automotive, meinl2017real}), the implementation of such protocols are feasible.} {\color{blue}To substantiate our claim, Table~\ref{tab:comp} reports the average execution time of a single evaluation of the Laplace transform of the interference distribution and the required summations, obtained using MATLAB, Python, and optimized C implementations. All tests were carried out on an Intel i7-1165G7, 2.8~GHz CPU with 16~GB RAM. The results indicate that our framework is executable within a fraction of the radar frame duration (10--50~ms), leaving sufficient computational margin. These results confirm that the proposed computations are well within real-time feasibility, even without hardware acceleration. Further improvements are possible using pre-computed lookup tables or FPGA/GPU offloading.}

{On centralized MAC:} Additionally, the proposed scheme consists of the derivation of an optimal ALOHA parameter which is to be calculated in the network operator for the connected cars and broadcast through network to vehicle (N2V) links. {Furthermore, the vehicle density across different times of the day and across different locations of a city albeit varying, remain relatively stable in the order of an hour. In this regard, similar to how a network operator sets the cell-range expansion parameters in a network of base stations, it is envisaged that the network operator will approximate the current vehicular density and fix the optimal ALOHA parameter for several hours. Indeed this is how most of macro network parameters are selected in current deployment of networks.} {\color{blue} From a systems implementation perspective, such {network-assisted parameter broadcasting} eliminates the instability associated with distributed contention-based optimization. Importantly, it allows the network to dynamically adapt ALOHA parameters to {spatio-temporal traffic variations} and {heterogeneous service requirements} across vehicles. The design is thus aligned with the ongoing {3GPP NR-V2X standardization efforts}, where the cellular network plays a key role in providing control-plane assistance for both communication and sensing functions~\cite{3gpp_rel18_v2x}.}

{Comparison to other MAC protocols:} {\color{blue}Existing MAC protocols, such as time-division multiplexing or contention-based schemes, primarily focus on coordinating medium access in a heuristic manner. While these methods are straightforward to implement, they lack analytical characterizations of performance under stochastic interference. In contrast, our proposed framework provides closed-form analytical guarantees on the tracking probability, offering insights into the fundamental limits of radar coexistence. A direct quantitative comparison is challenging because MAC-level heuristics and our analytical model address complementary aspects of the problem. Nevertheless, our results suggest that by integrating protocol-level scheduling with the interference-aware run-length analysis developed here, one could design MAC protocols that are both analytically tractable and practically efficient. We identify this integration as an important direction for future work.}

\section{Conclusion}
\label{sec:Con}
We introduced and characterized radar tracking probability, which is a novel metric to study automotive radar networks using stochastic geometry. Although the radar detection has been studied at length in literature, the statistics of successive successful detection has not been explored before. Using de Moivre's theorem, we derived the probability that an active radar experiences a success burst of length $\nu$ or more in a block of $T$ slots. The latency and reliability \ac{QoS} requirements of automotive use-cases can be mapped to the parameters $\nu$ and $T$ of our framework. We considered a block ALOHA based MAC protocol to control the set of active radars and demonstrated that optimal selection of block length is non-trivial. A larger block length improves the tracking performance of active radars in that block, but it results in a fewer radars getting activated within the radar deadline. For a given choice of block length, the optimal channel access parameter can be obtained via algorithms such as hill-climbing due to the unimodal nature of the tracking success with respect to the access parameter. In real world network, multiple use-cases co-exist in a vehicular network. An optimal frame design and access protocol for a network-wide utility function is an interesting direction of research which we will take up in a future work.

\appendices

\section{Proof of Lemma~\ref{le:lemma1}}
\label{app:lemma1}
\begin{figure}
    \centering
    \includegraphics[width=0.8\linewidth]{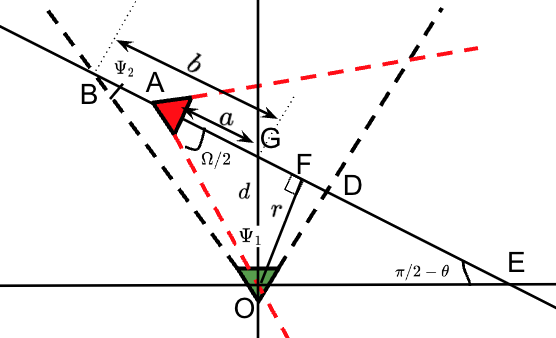}
    \caption{Evaluation of the interference region in a line characterized by $(r,\theta)$ intersecting the street containing the ego-radar (shown in red), i.e., the $y$-axis.}
    \label{fig:calculations}
\end{figure}

We derive the expressions of $a$ and $b$ using simple geometrical arguments from Fig.~\ref{fig:calculations}, {which corresponds to the scenario presented in Fig.~\ref{fig:illustration}. the ego-radar is shown in green at the origin $O$, while the interfering radar is located at $A$ and is shown in red. The interfering radar is present in an intersecting street $L_0$ which is is affine extension of the segment $BE$. The two key angles of interest are $\angle AOG = \psi_1$ and $\angle OBG = \psi_2$. Recall that the half-power beamwidth of the radar is $\Omega$. Hence the angle $\angle OAE = \frac{\Omega}{2}$. The interfering radar in this illustration is shown at a distance $AG = a$ from the intersection point. Now, employing the sine rule in the triangle $AOG$, we have:
$$\frac{a}{\sin \Psi_1} = \frac{d}{\sin \frac{\Omega}{2}}.$$
Now since the orientation of the intersecting line is $\theta$ with respect to the $x-$axis, $\angle OEA = \frac{\pi}{2} - \theta$. This implies $\angle OGE = \theta$ and accordingly, $\angle OGA = \pi - \theta$. Consequently, we see from the triangle OAG that $\psi_1 = \pi - \angle OAG  - \angle OGA = \pi - \frac{\Omega}{2} - (\pi - \theta) = \theta - \frac{\Omega}{2}$. Combining this with the sine rule, we have
\begin{align}
    a = \frac{d \sin \psi_1}{\sin \frac{\Omega}{2}} = \frac{d \sin \left(\theta - \frac{\Omega}{2}\right)}{\sin \frac{\Omega}{2}}. 
\end{align}
}
The value of $\Psi_2$ and accordingly $b$ follows similarly.

From sine rule, we have in triangle AOG, $\frac{a}{\sin \Psi_1} = \frac{d}{\sin \frac{\Omega}{2}}$, where $\Psi_1  = \angle AOG$. The value of $\Psi_1$ is derived by considering the triangle AOE, resulting in $\Psi_1 = \theta - \frac{\Omega}{2}$. 
The value of $\Psi_2$ and accordingly $b$ follows similarly.

\section{Proof of Lemma~\ref{lemma_csp}}
\label{app:csp}
{The conditional success probability is defined as the conditional CDF of $\xi(R)$, i.e.,
\begin{align}
    &P_{\rm d} = \mathbb{P} \left(\xi (R) \geq \gamma\; |\; \Phi\right) \\
    &= \mathbb{P}\left[\frac{\xi_0 {R}^{-2\alpha} \sigma_{\rm c}} {1 +  \xi_0 \sum_{{\bf w} \in \Phi^{\prime}} {h}_{\bf w} d_{\bf w}^{-\alpha}\textbf{1}({\bf w} \in \mathcal{C})} \geq \gamma\; |\; \Phi \right].
\end{align}
Here $\mathcal{C}$ denotes the active radar set in $\phi$ (a realization of $\Phi$). Now, due to the Swerling-I model for the fluctuating cross-section, $\sigma_{\rm c}$ is exponentially distributed with mean $\bar{\sigma}$. Accordingly, we calculate the above as
\begin{align}
    P_{\rm d} &=  \mathbb{P}\left[\sigma_{\rm c} \geq \frac{\gamma\left(1 +  \xi_0 \sum_{{\bf w} \in \Phi^{\prime}} {h}_{\bf w} d_{\bf w}^{-\alpha}\textbf{1}({\bf w} \in \mathcal{C})\right)}{\xi_0R^{-2\alpha}}\; |\; \Phi \right] \nonumber \\
    &\overset{(a)}{=} \mathbb{E}_{h_{\bf w}} \left[\exp\left(\frac{-\gamma - \gamma \xi_0 \sum_{{\bf w} \in \Phi^{\prime}} {h}_{\bf w} d_{\bf w}^{-\alpha} \textbf{1}({\bf w} \in \mathcal{C})} {\bar{\sigma}\xi_0 R^{-2\alpha}}\right)\right]. \nonumber
\end{align}
}
In the above, the step (a) is due to the fact that $\sigma_{\rm c}$ is exponentially distributed. Due to the conditioning on $\Phi$, the locations of the interferers, i.e., ${\bf w}$ is fixed and not random. The only randomness that remains in the resulting expression is due to the fast-fading $h_{\bf w}$, with respect to which we take the expectation in the remaining steps. This is evaluated as
{
\begin{align*}
 P_{\rm d}    &= e^{\left(\frac{-\gamma}{\bar{\sigma}\xi_0 R^{-2\alpha}}\right)} \mathbb{E}_{h_{\bf w}} \left[\exp\left(\frac{-\gamma \xi_0 \sum_{{\bf w} \in \Phi^{\prime}} {h}_{\bf w} d_{\bf w}^{-\alpha} \textbf{1}({\bf w} \in \mathcal{C})} {\bar{\sigma}\xi_0 R^{-2\alpha}}\right)\right]\\
    &= e^{\left(\frac{-\gamma}{\bar{\sigma}\xi_0 R^{-2\alpha}}\right)} \left(\prod_{{\bf w} \in \Phi^{\prime}} \mathbb{E}_{h_{\bf w}} \exp\left(\frac{-\gamma \xi_0 d_{\bf w}^{-\alpha} {h}_{w}} {\bar{\sigma}\xi_0 R^{-2\alpha}}\right)\right) \\
    &\overset{(b)}{=}  \underbrace{\exp{\left(\frac{-\gamma}{\xi_0 R^{-2\alpha}}\right)}}_{\text{Impact of noise}} \underbrace{\left( \prod_{{\bf w} \in \phi^{\prime}} \frac{1}{1+\frac{\gamma d_{\bf w}^{-\alpha}}{R^{-2\alpha}}} \right)}_{\text{Impact of interference}}.
\end{align*}
}
Step (b) follows from the Laplace transform of the exponentially distributed $h_{\bf w}$.

\section{Proof of Lemma~\ref{lem:moments}}
\label{app:momentscsp}
{
    From \eqref{eq:cond_succ} we can write
\begin{align}
    \mathbb{E}\left[P^l_{\rm d}\right] &= \mathbb{E}\left[ \exp{\left(\frac{-l\gamma}{\xi_0 R^{-2\alpha}}\right)} \left( \prod_{{\bf w} \in \Phi^{\prime}} \frac{1}{1+\frac{\gamma d_{\bf w}^{-\alpha}}{R^{-2\alpha}}} \right)^l\right] \nonumber \\
    &=  \exp{\left(\frac{-l\gamma}{\xi_0 R^{-2\alpha}}\right)} \mathbb{E}\left[\left( \prod_{{\bf w} \in \Phi^{\prime}} \frac{1}{1+\frac{\gamma d_{\bf w}^{-\alpha}}{R^{-2\alpha}}} \right)^l\right].
\end{align}
Now, the second term can be divided into two parts:
\begin{align}
    \mathbb{E}\left[\left( \prod_{{\bf w} \in \Phi^{\prime}} \frac{1}{1+\frac{\gamma d_{\bf w}^{-\alpha}}{R^{-2\alpha}}} \right)^l\right] = &\mathbb{E}\left[\left( \prod_{{\bf w} \in \Phi^{\prime}_0} \frac{1}{1+\frac{\gamma d_{\bf w}^{-\alpha}}{R^{-2\alpha}}} \right)^l\right]\cdot \nonumber \\
    &\mathbb{E}\left[\left( \prod_{{\bf w} \in \Phi^{\prime} \backslash \Phi^{\prime}_0} \frac{1}{1+\frac{\gamma d_{\bf w}^{-\alpha}}{R^{-2\alpha}}} \right)^l\right],
\end{align}
where $\Phi'_0$ is the thinned 1D \ac{PPP} on $L_0$. Now, thanks to Slivnyak's theorem, $\Phi' \backslash \Phi'_0$ has the same statistics as $\Phi'$. The first term is simply evaluated using the \ac{PGFL} of a 1D \ac{PPP} on $L_0$ with intensity $\delta \lambda$:
\begin{align}
    &\mathbb{E}\left[\left( \prod_{{\bf w} \in \Phi^{\prime}_0} \frac{1}{1+\frac{\gamma d_{\bf w}^{-\alpha}}{R^{-2\alpha}}} \right)^l\right] = \exp\left(- \lambda \delta \int_0^{\infty} 1 - \mathcal{T}_z^l{\rm d}z\right),
\end{align}
where $\mathcal{T}_z = \left(\frac{R^{-2\alpha}}{R^{-2\alpha} + \gamma z^{-\alpha}}\right)$. The second term, due to its doubly stochastic nature, has more involved form.
\begin{align}
    &\mathbb{E}\left[\left( \prod_{{\bf w} \in \Phi^{\prime} \backslash \Phi^{\prime}_0} \frac{1}{1+\frac{\gamma d_{\bf w}^{-\alpha}}{R^{-2\alpha}}} \right)^l\right] = \nonumber \\
    &\mathbb{E}_{\Phi_I}\left[\mathbb{E}_{\Phi'_i}\left[\left( \prod_{{\bf w} \in \Phi^{\prime}_i \backslash \Phi^{\prime}_0} \frac{1}{1+\frac{\gamma d_{\bf w}^{-\alpha}}{R^{-2\alpha}}} \right)^l\right] \bigg| \Phi_{I}\right] \nonumber\\
    &=\mathbb{E}_{\Phi_I}\left[\exp\left(-\delta\lambda \int_{a_i}^{b_i} 1 - \left(\frac{R^{-2\alpha}}{R^{-2\alpha}+\gamma w(x,z,\theta))^{-\alpha}}\right)^l {\rm d}z \right) \right] \nonumber \\
    & = \exp\left(-\lambda_L\int_0^{2\pi}\int_{0}^\infty 1 - \exp\left( -\delta\lambda\int_{a_i}^{b_i} 1 - \mathcal{T}^l_{z,x} {\rm d}z \right) \right. \nonumber\\
    & \hspace{3cm}\left.{\rm d}x {\rm d}\theta\right) \nonumber
\end{align}
where $w(x,z,\theta) = x^2 + z^2 + 2 x z \cos\theta$ and $\mathcal{T}_{z,x} = \frac{R^{-2\alpha}}{R^{-2\alpha}+\gamma w(x,z,\theta))^{-\alpha}}$.
}

\section{Proof of Theorem~\ref{theo:demoivre}}
\label{app:demoivre}
{Given that in each slot the detection success probability is $P_{\rm d}$, we are interested in the distribution of the run length of detection successes.  This problem is solved using generating functions. In particular, the combinatorial class of all sequences without a run of $m$ successes (S), i.e., $\xi \geq \gamma$  in a row can be written as
\[
\sum_{k\geq 0}(\mbox{length}_{< m }(S)\,F)^k \,\mbox{seq}_{< m }(S),
\]
where $F$ denotes the failure, i.e., $\xi < \gamma$, with corresponding counting generating function
\[
L(s,f)={\sum_{0\leq j< m }s^j\over 1-(\sum_{0\leq j< m }s^j)t}={1-s^ m \over 1-s-(1-s^ m )f}.
\]
We introduce the conditional success probability by replacing \( s \) with \( P_{\rm d} \) and \( f \) with \( q \), where \( q = 1 - P_{\rm d} \). The generating functional in terms of $s$ thus takes the form
\[
G(s)={1-P_{\rm d}^ m  s^ m \over1-s+P_{\rm d}^ m  s^{ m +1}(1 - P_{\rm d}))}.
\]
The coefficient of \( s^n \) in \( G(s) \) is $\mathbb{P}(L<m).$ Note that the function $1/(1-s(1-P_{\rm d}^ m  s^ m (1 - P_{\rm d}) ))$ can be rewritten as
\begin{align}
&\sum_{k\geq 0}s^k(1-P_{\rm d}^ m  s^ m (1 - P_{\rm d}) )^k \nonumber \\
&=\sum_{k\geq 0}\sum_{j\geq 0} {k\choose j} (-P_{\rm d}^ m (1 - P_{\rm d}))^js^{k+j m }.
\end{align}
The coefficient of $s^n$ in this function is $c(n)=\sum_{j\geq 0}{n-j m \choose j}(-P_{\rm d}^ m (1 - P_{\rm d}))^j$. Therefore the coefficient of $s^n$ in $G(s)$ is $c(n)-P_{\rm d}^ m  c(n- m )$. Finally,
\begin{align*} 
&\mathbb{P}(L\geq m)=1-\mathbb{P}(L<m)\\
=&P_{\rm d}^ m  c(n- m )+1-c(n)\\
=&P_{\rm d}^ m  \sum_{j\geq 0}(-1)^j{n-(j+1) m \choose j}(P_{\rm d}^ m (1 - P_{\rm d}))^j+ \\ &\sum_{j\geq 1}(-1)^{j+1}{n-j m \choose j}(P_{\rm d}^ m (1 - P_{\rm d}))^j\\
=&\sum_{j\geq 1}(-1)^{j+1}  \left[{n-j m \choose j-1}+{n-j m \choose j}(1 - P_{\rm d})\right] \\&P_{\rm d}^{ jm } (1 - P_{\rm d})^{j-1}\\
=&\sum_{j\geq 1}(-1)^{j+1}  \left[{n-j m \choose j-1}P_{\rm d}+{n-j m +1\choose j}(1 - P_{\rm d}) \right] \\&P_{\rm d}^{ jm} (1 - P_{\rm d})^{j-1}\\
=&\sum_{j\geq 1}(-1)^{j+1}  \left[P_{\rm d}+{n-j m +1\over  j}\, (1 - P_{\rm d})\right] \\& {n-j m \choose j-1}\,P_{\rm d}^{ jm} (1 - P_{\rm d})^{j-1}. 
\end{align*}
}
\section{Proof of Theorem~\ref{prop:burstproperty}}
\label{app:burst}
The average number of successful tracking events $B(k)$, in each of which the ego radar observes at least $v$ consecutive successful detection events is
\begin{equation*}
    \mathbb{E}[B(k) | \Phi]  = (T - \nu +1) P_{\rm d}^{\nu},
\end{equation*}
where  $P_{\rm d}^{v}$ is given by~\eqref{eq:cond_succ}. So, from \Cref{lem:moments}, the expected number of bursts averaged over the point process is
\begin{align*}
    \mathbb{E}[B(k)] &=  \mathbb{E}\left[\mathbb{E}[B | \Phi]\right] =  (T - L +1) \mathbb{E}\left[P_{\rm d}^L\right] \\
    &= (T - L +1) \zeta(L).
 \end{align*}
Similarly, to obtain the expected tracking length $\bar{L}$, we derive
\begin{align*}
    &\mathbb{E}[L] = \mathbb{E}\left[\sum_{l = 0}^T l P_{\rm d}^l (1 - P_{\rm d})\right]  \\
    &=\mathbb{E}\left[ \frac{T P_{\rm d}^{T+2} -TP_{\rm d}^{T+1}-P_{\rm d}^{T+1}+P_{\rm d}}{1-P_{\rm d}}\right]  \\
    &=\mathbb{E}\left[\frac{P_{\rm d} -P_{\rm d}^{T+1}}{1-P_{\rm d}}-TP_{\rm d}^{T+1}\right]  =\mathbb{E}\left[\sum_{l = 1}^{T} P_{\rm d}^l - TP^{T+1}_{\rm d}\right].
\end{align*}
Combining the last step with \Cref{lem:moments}, we obtain the expected burst length $\bar{L}$.

\bibliography{refer.bib}
\bibliographystyle{IEEEtran}

\end{document}